\documentclass[draftcls,  onecolumn]{IEEEtran}
\ifCLASSINFOpdf
\else
\fi
\usepackage{graphicx}
\usepackage{amsmath}
\usepackage{amssymb}
\usepackage{graphics}
\usepackage{latexsym}
\usepackage[dvips]{epsfig}
\usepackage[nospace]{cite}
\usepackage{subfigure}
\usepackage{setspace}
\usepackage{tabularx}
\usepackage{color}

\newtheorem{Definition}{Definition}
\newtheorem{Lemma}{Lemma}

\newtheorem{Proposition}[Lemma]{Proposition}
\newtheorem{Theorem}{Theorem}

\newtheorem{Remark}{Remark}

\usepackage[ colorlinks = true,
 linkcolor = blue,
 urlcolor = blue,
 citecolor = red,
 anchorcolor = green,
]{hyperref}

\allowdisplaybreaks[1]

\def\Pr{{\mathbb{P}}}
\def\E{{\mathbb{E}}}
\def\Var{{\rm {Var}}}

\begin{document}
%
\title{A Tight Upper Bound on the Second-Order Coding Rate of the Parallel Gaussian Channel with Feedback}
%
%
%


\author{Silas~L.~Fong and Vincent~Y.~F.~Tan
\thanks{S.~L.~Fong and V.~Y.~F.~Tan were supported by NUS Young Investigator Award under Grant R-263-000-B37-133.
 }%
\thanks{S.~L.~Fong is with the Department of Electrical and Computer Engineering, NUS, Singapore 117583 (e-mail: \texttt{silas\_fong@nus.edu.sg}).}%
\thanks{V.~Y.~F.~Tan is with the Department of Electrical and Computer Engineering, NUS, Singapore 117583, and also with the Department of Mathematics, NUS, Singapore 119076 (e-mail: \texttt{vtan@nus.edu.sg}).}}
\maketitle
\flushbottom

\begin{abstract}
This paper investigates the asymptotic expansion for the maximum rate of fixed-length codes over a parallel Gaussian channel with feedback under the following setting: A peak power constraint is imposed on every transmitted codeword, and the average error probabilities of decoding the transmitted message are non-vanishing as the blocklength increases.
The main contribution of this paper is a self-contained proof of an upper bound on the first- and second-order asymptotics of the parallel Gaussian channel with feedback. The proof techniques involve developing an information spectrum bound followed by using Curtiss' theorem to show that a sum of dependent random variables associated with the information spectrum bound converges in distribution to a sum of independent random variables, thus facilitating the use of the usual central limit theorem. Combined with existing achievability results, our result implies that the presence of feedback does not improve the first- and second-order asymptotics.
\end{abstract}


%
\IEEEpeerreviewmaketitle

\section{Introduction} \label{Introduction}
This paper considers a point-to-point communication scenario where a source wants to transmit a message to a destination through a set of independent additive white Gaussian noise (AWGN) channels. The set of independent AWGN channels is referred to as the \emph{parallel Gaussian channel} \cite[Sec.~9.4]{CoverBook} (also called the \emph{Gaussian product channel} in~\cite[Sec.~3.4.3]{elgamal}). The parallel Gaussian channel has been used to model
the multiple-input multiple-output (MIMO) channel~\cite[Sec.~7.1]{davidTseBook} --- an essential channel model in wireless communications. Suppose the parallel Gaussian channel consists of~$L$ independent AWGN channels, and
let $\mathcal{L}\stackrel{\text{def}}{=} \{1, 2, \ldots, L\}$ be the index set of the $L$~channels.
For the $k^{\text{th}}$ channel use, the relation for the $\ell^{\text{th}}$ channel between the input signal $X_{\ell,k}$ and output signal $Y_{\ell,k}$ is
\begin{equation}
Y_{\ell,k} = X_{\ell,k} + Z_{\ell,k} \label{channelLawFormulation1Intro}
\end{equation}
where $\{Z_{\ell,k}\}_{\ell \in \mathcal{L}}$ are independent Gaussian noises. For each $\ell\in \mathcal{L}$, the variance of the noise induced by the $\ell^{\text{th}}$~channel is assumed to be some positive number $N_\ell>0$ for all channel uses, i.e., $\Var[Z_{\ell,k}]=N_\ell$ for all $k\in \mathbb{N}$.
 To keep notation compact, let $\boldsymbol{X}_k$, $\boldsymbol{Y}_k$ and $\boldsymbol{Z}_k$ denote the random column vectors $[X_{1, k}\ X_{2,k}\ \ldots \ X_{L,k}]^t$, $[Y_{1,k}\ Y_{2,k}\ \ldots \ Y_{L,k}]^t$ and $[Z_{1,k}\ Z_{2,k}\ \ldots \ Z_{L,k}]^t$ respectively. Then, the channel law~\eqref{channelLawFormulation1Intro} can be written as
\begin{equation}
\boldsymbol{Y}_k = \boldsymbol{X}_k + \boldsymbol{Z}_k. \label{channelLaw}
 \end{equation}
Throughout this paper, we consider fixed-length codes over the parallel Gaussian channel, where the block length is denoted by~$n$ unless specified otherwise. Every codeword $\boldsymbol{X}^n$ transmitted by the source over~$n$ channel uses is subject to the following \emph{peak power constraint} where $P>0$ denotes the permissible power for~$\boldsymbol{X}^n$:
\begin{align}
\Pr\left\{\frac{1}{n}\sum_{\ell=1}^L \sum_{k=1}^n X_{\ell,k}^2 \le P\right\}=1.  \label{peakPowerConstraint}
\end{align}
 \indent
If we would like to transmit a uniformly distributed message $W\in \{1, 2, \ldots, \lceil 2^{nR} \rceil\}$ over this channel where the error probabilities are required to vanish as the blocklength~$n$ approaches infinity, it was shown by Shannon~\cite{Sha49} that the maximum rate of communication~$R$ converges to a certain limit called {\em capacity}. The closed-form expression of the capacity can be obtained by finding the optimal power allocation among the~$L$ channels, which is described as follows.
Define the mapping $\mathrm{C}(\mathbf{s}) : \mathbb{R}_+^L \rightarrow \mathbb{R}_+$ as
\begin{equation}
\mathrm{C}(\mathbf{s})= \sum_{\ell=1}^L \frac{1}{2}\log\left(1+\frac{s_\ell}{N_\ell}\right) \label{defCPparallelNFB}
\end{equation}
where $s_\ell$ can be viewed as the power allocated to channel~$\ell$.
 If we let $\Lambda$, $P_1$, $P_2$, $\ldots$, $P_L$ denote the $L+1$ real numbers yielded from the water-filling algorithm~\cite[Ch~9.4]{CoverBook} where
\begin{equation}
\sum_{\ell=1}^L P_\ell = P \label{sumPell=P}
\end{equation}
and
\begin{equation}
P_\ell = \max\{0,\Lambda - N_\ell\} \label{pEllValue}
\end{equation}
for each $\ell\in \mathcal{L}$ and let
\begin{equation}
\mathbf{P}^*\stackrel{\text{def}}{=}[P_1\ P_2\ \ldots \ P_L]^t \label{pEllValue*}
\end{equation}
be the optimal power allocation vector,
 then the capacity of the parallel Gaussian channel was shown in~\cite{Sha49} to be $\mathrm{C}(\mathbf{P}^*)$ bits per channel use.
%
%
%
%
More specifically, if $M^*(n,\varepsilon,P)$ denotes the maximum number of messages that can be transmitted over $n$ channel uses with permissible power~$P$ and average error probability $\varepsilon$, one has
\begin{equation}
\lim_{\varepsilon\to 0}\liminf_{n\to\infty}\frac{1}{n}\log M^*(n,\varepsilon,P)=\mathrm{C}(\mathbf{P}^*). \label{shannonCapacity}
\end{equation}
The capacity result~\eqref{shannonCapacity} has been strengthened by Polyanskiy-Poor-Verd\'u \cite[Th.~78]{Pol10} and Tan-Tomamichel \cite[Appendix~A]{TanTom13a} for each $\varepsilon\in(0,1)$ as
 \begin{equation}
\frac{1}{n}\log M^*(n,\varepsilon,P)=\mathrm{C}(\mathbf{P}^*) + \sqrt{\frac{\mathrm{V}(\mathbf{P}^*)}{n}}\, \Phi^{-1} (\varepsilon) + \Theta\Big(\frac{\log n}{n}\Big), \label{eqn:asymp_expans}
\end{equation}
where $\mathrm{V} : \mathbb{R}_+^L \rightarrow \mathbb{R}_+$ is
the Gaussian dispersion function defined as
\begin{equation}
\mathrm{V}(\mathbf{s})=   \sum_{\ell=1}^L \frac{\frac{s_\ell}{N_\ell}(\frac{s_\ell}{N_\ell}+2) }{2(\frac{s_\ell}{N_\ell}+1)^2} \label{defVP}
\end{equation}
and $\Phi$ is the cumulative distribution function (cdf) of the standard normal distribution.
%

{\em Feedback}, which is the focus of the current paper, can simplify coding schemes and improve the performance of communication systems in many scenarios. See \cite[Ch.~17]{elgamal} for a thorough discussion on the benefits of feedback in single- and multi-user information theory. When feedback is allowed, each input symbol $\boldsymbol{X}_k$ depends on not only the transmitted message $W$ but also all the previous channel outputs up to the $(k-1)^{\text{th}}$ channel use, i.e., the symbols $(\boldsymbol{Y}_1, \boldsymbol{Y}_2, \ldots, \boldsymbol{Y}_{k-1})$. 
In the presence of noiseless feedback, let $M_{\text{fb}}^*(n,\varepsilon,P)$ denote the maximum number of messages that can be transmitted over $n$ channel uses with permissible power~$P$ and average error probability $\varepsilon$.
It was shown by Shannon~\cite{Sha56} that the presence of noiseless feedback does not increase the capacity of point-to-point \emph{memoryless channels}, which together with~\eqref{shannonCapacity} implies that
\begin{align}
\lim_{\varepsilon\to 0}\liminf_{n\to\infty}\frac{1}{n}\log M_{\text{fb}}^*(n,\varepsilon,P)=\mathrm{C}(\mathbf{P}^*). \label{shannonFBCapacity}
\end{align}
In view of~\eqref{eqn:asymp_expans}, we conclude that
\begin{equation}
\frac{1}{n}\log M_{\text{fb}}^*(n,\varepsilon,P)\ge \mathrm{C}(\mathbf{P}^*) + \sqrt{\frac{\mathrm{V}(\mathbf{P}^*)}{n}}\, \Phi^{-1} (\varepsilon) + \Theta\Big(\frac{\log n}{n}\Big). \label{eqn:asymp_expans*}
\end{equation}
In this paper, the main contribution is a conceptually simple, concise and self-contained proof that in the presence of feedback, the first- and second-order terms in the asymptotic expansion in~\eqref{eqn:asymp_expans} remains unchanged, i.e.,
\begin{equation}
\frac{1}{n}\log M_{\mathrm{fb}}^*(n,\varepsilon,P)=\mathrm{C}(\mathbf{P}^*) + \sqrt{\frac{\mathrm{V}(\mathbf{P}^*)}{n}}\, \Phi^{-1} (\varepsilon) + o\Big(\frac{1}{\sqrt{n}}\Big).\label{eqn:asymp_expans_fb}
\end{equation}



\subsection{Related Work} \label{subsecRelatedWork}
Our work is inspired by the recent study of the fundamental limits of communication over discrete memoryless channels (DMCs) with feedback~\cite{AW14}. It was shown by Altu\u{g} and Wagner~\cite[Th.~1]{AW14} that for some classes of DMCs whose capacity-achieving input distributions are not unique (in particular, the minimum and maximum conditional information variances differ), coding schemes with feedback achieve a better second-order asymptotics compared to those without feedback. They also showed \cite[Th.~2]{AW14} that feedback does not improve the second-order asymptotics of DMCs $q_{Y|X}$ if the conditional variance of the log-likelihood ratio $\log \frac{q_{Y|X}(Y|x)}{p^*(Y)}$, where $p^*$ is the unique capacity-achieving output distribution, does not depend on the input $x$. Such DMCs include the class of weakly-input symmetric DMCs initially studied by Polyanskiy-Poor-Verd\'u~\cite{PPV11b}.

However, we note that the proof technique used by Altu\u{g} and Wagner requires the use of a Berry-Ess\'een-type result for bounded martingale difference sequences~\cite{Machkouri}, and their technique cannot be extended to the parallel Gaussian channel with feedback because each input symbol~$X_{\ell,k}$ belongs to an interval $[-\sqrt{nP}, \sqrt{nP}]$ that grows without bound as~$n$ increases. Instead, our proof uses Curtiss' theorem to show that a sum of dependent random variables that naturally appears in the non-asymptotic analysis converges in distribution to a sum of independent random variables, thus facilitating the use of the usual central limit theorem~\cite{feller}.  

For $L=1$, the parallel Gaussian channel with feedback reduces to the AWGN channel with feedback, whose second-order coding rate is identical to the same channel without feedback by the following symmetry argument: The log-likelihood ratios $\log \frac{q_{Y|X}(Y|x)}{p^*(Y)}$ for all $x$ on the power sphere with radius $\sqrt{nP}$ are the same. See~\cite{FongTan15} for a rigorous but simple proof. In contrast, for $L>1$, this symmetry argument no longer holds due to the flexible power allocation among the~$L$ channels, and hence the simple proof suggested in~\cite{FongTan15} cannot be extended to the parallel Gaussian channel with feedback.

If the peak power constraint in~\eqref{peakPowerConstraint} is replaced with the expected power constraint $\E\left[\frac{1}{n}\sum_{\ell=1}^L \sum_{k=1}^n X_{\ell,k}^2 \right]\le P$, the first-order coding rate of the AWGN channel with feedback is improved from $\mathrm{C}(P)$ to $\mathrm{C}(\frac{P}{1-\varepsilon})$~\cite[Sec.~II]{TFT17} (the exact improvement holds for the non-feedback case as well~\cite[Sec.\ 4.3.3]{Pol10}) where $\varepsilon$ denotes the tolerable error probability. For the general case $L>1$, the proof in \cite[Sec.~II]{TFT17} can be easily extended to show that the first-order coding rate of the parallel Gaussian channel with feedback can be improved from $\mathrm{C}(\mathbf{P}^*)$ to $\mathrm{C}(\frac{\mathbf{P}^*}{1-\varepsilon})$, and hence~\eqref{eqn:asymp_expans_fb} no longer holds.

\subsection{Paper Outline}
This paper is organized as follows. The next subsection summarizes the notation used in this paper. Section~\ref{sectionDefinition} provides the problem setup of the parallel Gaussian channel with feedback under the peak power constraint and presents our main theorem. Section~\ref{sectionPrelim} contains the preliminaries required for the proof of our main theorem. The preliminaries include the following: (i) Important properties of non-asymptotic binary hypothesis testing quantities; (ii) Modification of power allocation among the parallel channels; (iii) Curtiss' theorem. Section~\ref{sectionMainResult} presents the proof of our main theorem.
Section~\ref{sectionConclusion} concludes this paper by explaining the novel ingredients in the proof of the main theorem and the major difficulty in strengthening the main theorem.
\subsection{Notation}\label{notation}
The sets of natural numbers, non-negative integers, real numbers and non-negative real numbers are denoted by $\mathbb{N}$, $\mathbb{Z}_+$, $\mathbb{R}$ and $\mathbb{R}_+$ respectively.
 An $L$-dimensional column vector is denoted by $\mathbf{a} \stackrel{\text{def}}{=} [a_1\ a_2\ \ldots \ a_L]^t$ where $a_\ell$ denote the $\ell^{\text{th}}$ element of $\mathbf{a}$. The Euclidean norm of a vector $\mathbf{a}\in \mathbb{R}^L$ is denoted by 
$\|\mathbf{a}\|_{2}\stackrel{\text{def}}{=} \sqrt{\sum_{\ell=1}^L a_\ell^2}$\,. We will take all logarithms to base~$e$ throughout this paper.

We use $\Pr\{\mathcal{E}\}$ to represent the probability of an
event~$\mathcal{E}$, and we let $\mathbf{1}\{\mathcal{E}\}$ be the indicator function of $\mathcal{E}$. Every random variable is denoted by a capital letter (e.g., $X$), and the realization and the alphabet of the random variable are denoted by the corresponding small letter (e.g., $x$) and calligraphic letter (e.g., $\mathcal{X}$) respectively.
We use $X^n$ to denote a random tuple $(X_1,  X_2,  \ldots ,  X_n)$, where all the elements $X_k$ have the same alphabet~$\mathcal{X}$. We let $p_X$ be the probability distribution of a random variable $X$. More specifically, $p_X$ is the Radon-Nikodym derivative of a measure with respect to the Lebesgue measure in an appropriate Euclidean space. We let $p_{Y|X}$ denote the conditional probability distribution of $Y$ given $X$ for any random variables~$X$ and~$Y$.
We let $p_Xp_{Y|X}$ denote the joint distribution of $(X,Y)$, i.e., $p_Xp_{Y|X}(x,y)=p_X(x)p_{Y|X}(y|x)$ for all $x$ and $y$.
 For any random variable~$X\sim p_X$ and any real-valued function~$g$ whose domain includes $\mathcal{X}$, we let $\Pr_{p_X}\{g(X)\ge\xi\}$ denote $\int_{\mathcal{X}} p_X(x)\mathbf{1}\{g(x)\ge\xi\}\, \mathrm{d}x$ for any real constant $\xi$ where $p_X$  The expectation and the variance of~$g(X)$ are denoted as
$
\E_{p_X}[g(X)]$ and
$
 \Var_{p_X}[g(X)]$ respectively.
 For simplicity, we drop the subscript of a notation if there is no ambiguity.
 For any real-valued Gaussian random variable $Z$ whose mean and variance are $\mu$ and $\sigma^2$ respectively, we let
 \begin{equation}
 \mathcal{N}(z; \mu, \sigma^2)\stackrel{\text{def}}{=} \frac{1}{\sqrt{2\pi \sigma^2}}e^{-\frac{(z-\mu)^2}{2\sigma^2}} \label{normalDist}
 \end{equation}
 be the corresponding probability density function.

\section{Parallel Gaussian Channel with Feedback} \label{sectionDefinition}
 Let~$\mathrm{s}$ and~$\mathrm{d}$ denote the source and the destination respectively. Suppose node~$\mathrm{s}$ transmits a message to node~$\mathrm{d}$ over $n$ channel uses through the $L$~independent AWGN channels.
 Before any transmission begins, node~$\mathrm{s}$ chooses message~$W$ destined for node~$\mathrm{d}$ where $W$ is uniformly distributed on the message alphabet
\begin{equation}
\mathcal{W}\stackrel{\text{def}}{=} \{1, 2, \ldots, M\} \label{defMessageAlphabet}
 \end{equation}
whose size is denoted by $M$. 
For the $k^{\text{th}}$ channel use, node~$\mathrm{s}$ transmits $\boldsymbol{X}_{k}$ and the corresponding channel output~$\boldsymbol{Y}_{k}$ satisfies~\eqref{channelLaw}.
 We assume that a noiseless feedback link from the destination node $\rm d$ to the source node $\rm s$ exists so that $(W, \boldsymbol{Y}^{k-1})$ is available for encoding $\boldsymbol{X}_k$ for each $k\in\{1, 2, \ldots, n\}$. In addition, the codewords $\boldsymbol{X}^n$ transmitted by~$\mathrm{s}$ is subject to the peak power constraint~\eqref{peakPowerConstraint}. Upon receiving $\boldsymbol{Y}^n$, node~$\mathrm{d}$ declares $\hat W$ to be the transmitted message.
\medskip
\begin{Definition} \label{defCode}
An $(n, M, P)$-feedback code consists of the
following:
\begin{enumerate}
\item A message set $\mathcal{W}$ at node~$\mathrm{s}$ as defined in~\eqref{defMessageAlphabet}. Message $W$ is uniform on $\mathcal{W}$.

\item An encoding function
\[
f_{\ell,k} : \mathcal{W}\times \mathbb{R}^{L\times (k-1)}\rightarrow \mathbb{R}
 \]
 for each $\ell\in\mathcal{L}$ and each $k\in\{1, 2, \ldots, n\}$, where $f_{\ell,k}$ is the encoding function at node~$\mathrm{s}$ for encoding $X_{\ell,k}$ such that
\[
X_{\ell,k}=f_{\ell,k} (W, \boldsymbol{Y}^{k-1})
\]
and the peak power constraint~\eqref{peakPowerConstraint} holds.
\item A decoding function
\[
\varphi :
\mathbb{R}^{L\times n} \rightarrow \mathcal{W},
 \]
where $\varphi$ is the decoding function for $W$ at node~$  \mathrm{d}$ such that
 \[
 \hat W = \varphi(\boldsymbol{Y}^{n}).
 \]
\end{enumerate}
\end{Definition}
\medskip
\begin{Definition}\label{defAWGNchannel}
 Let $\boldsymbol{X}$ and $\boldsymbol{Y}$ denote the random vectors $[X_{1}\ X_{2}\ \ldots \ X_{L}]^t$ and $[Y_{1}\ Y_{2}\ \ldots \ Y_{L}]^t$ respectively, and let $\mathbf{x}$ and $\mathbf{y}$ be their realizations respectively. The \emph{parallel Gaussian channel with feedback} is characterized by the conditional probability density distribution $q_{\boldsymbol{Y}|\boldsymbol{X}}$ satisfying
\begin{equation}
q_{\boldsymbol{Y}|\boldsymbol{X}}(\mathbf{y}|\mathbf{x})= \prod_{\ell=1}^L \mathcal{N}(y_\ell; x_\ell, N_\ell) \label{defAWGNchannelParallel}
\end{equation}
such that the following holds for any $(n, M, P)$-feedback code: For each $k\in\{1, 2, \ldots, n\}$,
\begin{align}
p_{W, \boldsymbol{X}^k, \boldsymbol{Y}^k}
 = p_{W, \boldsymbol{X}^k, \boldsymbol{Y}^{k-1}}p_{\boldsymbol{Y}_k|\boldsymbol{X}_k} \label{memorylessStatement*}
\end{align}
where
\begin{equation}
p_{\boldsymbol{Y}_k|\boldsymbol{X}_k}(\mathbf{y}_k|\mathbf{x}_k) = q_{\boldsymbol{Y}|\boldsymbol{X}}(\mathbf{y}_k|\mathbf{x}_k) \label{defChannelInDefinition*}
\end{equation}
for all $(\mathbf{x}^n,\mathbf{y}^n)\in \mathbb{R}^{L\times n}\times\mathbb{R}^{L\times n}$.
\end{Definition}
\medskip

 For any $(n, M, P)$-feedback code, let $p_{W,\boldsymbol{X}^n, \boldsymbol{Y}^n, \hat W}$ be the joint distribution induced by the code. We can use Definition~\ref{defCode}, \eqref{memorylessStatement*} and \eqref{defChannelInDefinition*} to factorize $p_{W,\boldsymbol{X}^n, \boldsymbol{Y}^n, \hat W}$ as follows:
\begin{align}
 p_{W,\boldsymbol{X}^n, \boldsymbol{Y}^n, \hat W}
& = p_W \left(\prod_{k=1}^n p_{\boldsymbol{X}_k|W, \boldsymbol{Y}^{k-1}} p_{\boldsymbol{Y}_k|\boldsymbol{X}_k}\right)p_{\hat W |\boldsymbol{Y}^n}. \label{memorylessStatement}
\end{align}
\medskip
\begin{Definition} \label{defErrorProbability}
For an $(n, M, P)$-feedback code, we can calculate according to~\eqref{memorylessStatement} the \textit{average probability of decoding error} defined as $\Pr\big\{\hat W \ne W\big\}$.
We call an $(n, M, P)$-feedback code with average probability of decoding error no larger than $\varepsilon$ an $(n, M, P, \varepsilon)$-feedback code.
\end{Definition}
\medskip
Define
\[
M_{\mathrm{fb}}^*(n, \varepsilon, P) \stackrel{\text{def}}{=} \max\left\{M\in \mathbb{N} \left|\: \text{There exists an $(n, M, P, \varepsilon)$-feedback code}\right.\right\}.
\]
\begin{Definition}\label{defCapacity}
Let $\varepsilon\in (0,1)$. The $\varepsilon$-capacity of the parallel Gaussian channel with feedback, denoted by $C_\varepsilon^{\text{fb}}$, is defined to be
\[
C_\varepsilon^{\text{fb}} \stackrel{\text{def}}{=} \liminf\limits_{n\rightarrow \infty}\frac{1}{n}\log M_{\mathrm{fb}}^*(n, \varepsilon, P).
\]
The capacity is defined to be
\[
C_0^{\text{fb}}\stackrel{\text{def}}{=} \inf_{\varepsilon > 0} C_\varepsilon^{\text{fb}}.
\]
\end{Definition}

\begin{Definition} \label{defDispersion}
Let $\varepsilon\in (0,1)$.  The $\varepsilon$-second-order coding rate of the parallel Gaussian channel with feedback, denoted by $\mathrm{L}_\varepsilon^{\text{fb}}$, is defined to be
\[
\mathrm{L}_\varepsilon^{\text{fb}} \stackrel{\text{def}}{=} \liminf\limits_{n\rightarrow \infty}\frac{1}{\sqrt{n}}\left(\log M_{\mathrm{fb}}^*(n, \varepsilon, P)-nC_\varepsilon^{\text{fb}}\right).
\]
\end{Definition}

Recall how $\mathrm{C}(\mathbf{P}^*)$ and $\mathrm{V}(\mathbf{P}^*)$ are determined through \eqref{defCPparallelNFB}, \eqref{sumPell=P}, \eqref{pEllValue}, \eqref{pEllValue*} and \eqref{defVP}.
Since the capacity of the parallel Gaussian channel without feedback is $\mathrm{C}(\mathbf{P}^*)$ (see, e.g., \cite{Sha49} and \cite[Sec.\ 3.4.3]{elgamal}) and an introduction of an extra noiseless feedback link does not increase the capacity~(see, e.g., \cite{Sha56} and \cite[Sec.~9.6]{CoverBook}), it follows that
\begin{equation}
C_0^{\text{fb}}=\mathrm{C}(\mathbf{P}^*). \label{wellKnownFBCapacity}
\end{equation}
Before stating our main result, recall that $\Phi: (-\infty, \infty)\rightarrow (0,1)$ is the cdf of the standard normal distribution. Since $\Phi$ is strictly increasing on $(-\infty, \infty)$, the inverse of $\Phi$ is well-defined and is denoted by $\Phi^{-1}$. The following theorem is the main result in this paper.
\medskip
\begin{Theorem} \label{thmMainResult}
Fix an $\varepsilon \in (0,1)$.
Then,
\begin{equation}
C_\varepsilon^{\text{fb}} = \mathrm{C}(\mathbf{P}^*)
\end{equation}
and the $\varepsilon$-second-order coding rate satisfies
\begin{equation}
\mathrm{L}_\varepsilon^{\text{fb}} \le \sqrt{\mathrm{V}(\mathbf{P}^*)}\, \Phi^{-1}(\varepsilon).
\end{equation}
\end{Theorem}
\medskip

Combining \eqref{eqn:asymp_expans} and Theorem~\ref{thmMainResult}, we complete the characterizations of the first- and second-order asymptotics of the parallel Gaussian channel with feedback as shown in~\eqref{eqn:asymp_expans_fb}.


\section{Preliminaries for the Proof of Theorem \ref{thmMainResult}} \label{sectionPrelim}
\subsection{Binary Hypothesis Testing} \label{sectionBHT}
The following definition concerning the non-asymptotic fundamental limits of a simple binary hypothesis test is standard. See for example  \cite[Section~2.3]{Pol10}.
\medskip
\begin{Definition}\label{defBHTDivergence}
Let $p_{X}$ and $q_{X}$ be two probability distributions on some common alphabet $\mathcal{X}$. Let
\[
\mathcal{A}(\{0,1\}|\mathcal{X})\stackrel{\text{def}}{=} \{
r_{Z|X}: \text{$Z$ and $X$ assume values in $\{0,1\}$ and $\mathcal{X}$ respectively}\}
\]
be the set of randomized binary hypothesis tests between $p_{X}$ and $q_{X}$ where $\{Z=0\}$ indicates the test chooses $q_X$, and let $\delta\in [0,1]$ be a real number. The minimum type-II error in a simple binary hypothesis test between $p_{X}$ and $q_{X}$ with type-I error less than $1-\delta$ is defined as
\begin{align}
 \beta_{\delta}(p_X\|q_X) \stackrel{\text{def}}{=}
\inf\limits_{\substack{r_{Z|X} \in \mathcal{A}(\{0,1\}|\mathcal{X}): \\ \int_{\mathcal{X}}r_{Z|X}(1|x)p_X(x)\, \mathrm{d}x\ge \delta}} \int_{\mathcal{X}}r_{Z|X}(1|x)q_X(x)\, \mathrm{d}x.\label{eqDefISDivergence}
\end{align}
\end{Definition}\medskip
The existence of a minimizing test $r_{Z|X}$ is guaranteed by the Neyman-Pearson lemma.

We state in the following lemma and proposition some important properties of $\beta_{\delta}(p_X\|q_X)$, which are crucial for the proof of Theorem~\ref{thmMainResult}. The proof of the following lemma can be found  in, for example, \cite[Lemma~1]{Wang09}.
\medskip
\begin{Lemma}\label{lemmaDPI} Let $p_{X}$ and $q_{X}$ be two probability distributions on some $\mathcal{X}$, and let $g$ be a function whose domain contains $\mathcal{X}$. Then, the following two statements hold:
\begin{enumerate}
\item[1.] (Data processing inequality (DPI)) $\beta_{\delta}(p_X\|q_X) \le \beta_{\delta}(p_{g(X)}\|q_{g(X)})$.
\item[2.] For all $\xi>0$, $\beta_{\delta}(p_X\|q_X)\ge \frac{1}{\xi}\left(\delta - \int_{\mathcal{X}}p_X(x) \boldsymbol{1}\left\{ \frac{p_X(x)}{q_X(x)} \ge \xi \right\}\, \mathrm{d}x\right) $.
\end{enumerate}
\end{Lemma}
\medskip
 The proof of the following proposition can be found in~\cite[Lemma~3]{Wang09} (see also~\cite[Th.~27]{PPV10}).
 \medskip
\begin{Proposition} \label{propositionBHTLowerBound}
Let $p_{U,V}$ be a probability distribution defined on $\mathcal{W}\times \mathcal{W}$ for some finite alphabet $\mathcal{W}$, and let $p_U$ be the marginal distribution of $p_{U,V}$. In addition, let $q_{V}$ be a distribution defined on $\mathcal{W}$. Suppose $p_{U}$ is the uniform distribution, and let
\begin{equation}
\alpha = \Pr\{U\ne V\} \label{defAlpha}
\end{equation}
be a real number in $[0, 1]$ where $(U,V)$ is distributed according to $p_{U,V}$. Then,
\begin{equation}
|\mathcal{W}| \le 1/\beta_{1-\alpha}(p_{U,V}\|p_{U} q_{V}). \label{propositionBHTLowerBoundEq1}
\end{equation}
\end{Proposition}
\subsection{Modification of Power Allocation among the Parallel Channels}
For each transmitted codeword $\mathbf{x}^n\in \mathbb{R}^{L\times n}$, we can view $\sum_{k=1}^n x_{\ell,k}^2$ as the power allocated to the $\ell^{\text{th}}$ channel for each $\ell\in\mathcal{L}$. In the proof of Theorem~\ref{thmMainResult}, an early step is to discretize the power allocated to the $L$ channels. To this end, we need the following definition which defines the power allocation vector of a sequence $\mathbf{x}^n\in \mathbb{R}^{L\times n}$.
\medskip
\begin{Definition} \label{definitionPowerType}
The \emph{power allocation mapping} $\phi : \mathbb{R}^{L\times n} \rightarrow \mathbb{R}_+^L$ is defined as
\begin{equation}
\phi(\mathbf{x}^n)=\frac{1}{n}\left[\sum_{k=1}^n x_{1,k}^2\ \sum_{k=1}^n x_{2,k}^2\ \ldots \ \sum_{k=1}^n x_{L,k}^2\right]^t.
\end{equation}
We call $\phi(\mathbf{x}^n)$ the \emph{power type of $\mathbf{x}^n$}.
\end{Definition}

\medskip
The proof of Theorem~\ref{thmMainResult} involves modifying a given length-$n$ code so that useful bounds on the performance of the given code can be obtained by analyzing the modified code. More specifically, the encoding functions the given code are modified so that the power type of the random codeword generated by the modified code always falls into some small bounding box. The specific modification of the encoding functions is described in the following definition.
\medskip
\begin{Definition} \label{definitionTransformedCode}
Given an $(n, M, P)$-feedback code, let $\mathcal{W}$, $\{f_{\ell,k}|1\le \ell\le L, 1\le k\le n\}$ and $\varphi$ be the corresponding message alphabet, encoding functions and decoding function respectively. In addition, let $\gamma \ge 0$ and $\mathbf{s}=[s_1\ s_2\ \ldots \ s_L]\in \mathbb{R}_+^L$ such that $\sum_{\ell=1}^L s_\ell = P$. Then, the \emph{$(\gamma, \mathbf{s})$-modified code based on the $(n, M, P)$-feedback code} consists of the following message alphabet, encoding functions and decoding function which are denoted by $\tilde{\mathcal{W}}$, $\{\tilde f_{\ell,k}|1\le \ell\le L, 1\le k\le n\}$ and $\tilde \varphi$ respectively:
\\
\textbf{1)}  A message set $\tilde{\mathcal{W}}=\mathcal{W}$ at node~$\mathrm{s}$. Message $W$ is uniform on $\tilde{\mathcal{W}}$.
\\
\textbf{2)} An encoding function
\[
\tilde f_{\ell,k} : \mathcal{W}\times \mathbb{R}^{L\times (k-1)}\rightarrow \mathbb{R}
 \]
 for each $\ell\in\mathcal{L}$ and each $k\in\{1, 2, \ldots, n\}$, which is defined as follows. For each $w\in \mathcal{W}$ and each $\mathbf{y}^{k-1}\in  \mathbb{R}^{L\times (k-1)}$, define $\tilde f_{\ell,k}$ in a recursive manner in this order
$
 \tilde f_{1,1}, \tilde f_{2, 1}, \ldots, \tilde f_{L, 1}, \ldots,  \tilde f_{1,n}, \tilde f_{2, n}, \ldots, \tilde f_{L, n}
$
 as follows: For each $k=1, 2, \ldots, n-1$, define $\tilde f_{\ell,k}$ recursively for $\ell=1, 2, \ldots, L$ as
 \begin{align}
\tilde f_{\ell,k}(w, \mathbf{y}^{k-1})=\begin{cases}
f_{\ell,k}(w, \mathbf{y}^{k-1}) & \parbox[t]{4 in}{if $f_{\ell,k}(w, \mathbf{y}^{k-1})^2+\sum\limits_{i=1}^{k-1} \tilde f_{\ell,i}(w, \mathbf{y}^{i-1})^2 \le n(s_\ell+\gamma)$,
}\\ 0 & \parbox[t]{4 in}{if $f_{\ell,k}(w, \mathbf{y}^{k-1})^2+\sum\limits_{i=1}^{k-1} \tilde f_{\ell,i}(w, \mathbf{y}^{i-1})^2 > n(s_\ell+\gamma)$.
}
 \end{cases} \label{defTildeFeq1}
\end{align}
It follows from~\eqref{defTildeFeq1} that
\begin{equation}
\Pr\left\{\bigcap_{\ell=1}^L\bigg\{\sum\limits_{i=1}^{n-1} \tilde f_{\ell,i}(W, \boldsymbol{Y}^{i-1})^2 \le n(s_\ell+\gamma)\bigg\}\right\}=1 \label{defTildeFeq1*}
\end{equation}
and
\begin{equation}
\Pr\left\{\sum\limits_{\ell=1}^{L}\sum\limits_{i=1}^{n-1} \tilde f_{\ell,i}(W, \boldsymbol{Y}^{i-1})^2 \le \sum\limits_{\ell=1}^{L}\sum\limits_{i=1}^{n-1} f_{\ell,i}(W, \boldsymbol{Y}^{i-1})^2\right\}=1. \label{defTildeFeq1**}
\end{equation}
 In addition, in view of~\eqref{defTildeFeq1*}, we define $\tilde f_{\ell,n}$ recursively for $\ell=1, 2, \ldots, L-1$ as follows:
 \begin{align}
&\tilde f_{\ell,n}(w, \mathbf{y}^{n-1})\notag\\*
&=\begin{cases}
f_{\ell,n}(w, \mathbf{y}^{n-1}) & \parbox[t]{5 in}{if $f_{\ell,n}(w, \mathbf{y}^{n-1})^2+\sum\limits_{i=1}^{n-1} \tilde f_{\ell,i}(w, \mathbf{y}^{i-1})^2 \in[n(s_\ell-L\gamma), n(s_\ell+\gamma)]$,}\\
 \sqrt{n(s_\ell-L\gamma) - \sum\limits_{i=1}^{n-1} \tilde f_{\ell,i}(w, \mathbf{y}^{i-1})^2}& \parbox[t]{5 in}{if $f_{\ell,n}(w, \mathbf{y}^{n-1})^2+\sum\limits_{i=1}^{n-1} \tilde f_{\ell,i}(w, \mathbf{y}^{i-1})^2 < n(s_\ell-L\gamma)$,}
 \\
\sqrt{n(s_\ell+\gamma) - \sum\limits_{i=1}^{n-1} \tilde f_{\ell,i}(w, \mathbf{y}^{i-1})^2} &   \parbox[t]{5 in}{if $f_{\ell,n}(w, \mathbf{y}^{n-1})^2+\sum\limits_{i=1}^{n-1} \tilde f_{\ell,i}(w, \mathbf{y}^{i-1})^2 > n(s_\ell-L\gamma)$.} 
 \end{cases} \label{defTildeFeq2}
\end{align}
Combining~\eqref{defTildeFeq1} and \eqref{defTildeFeq2}, we conclude that
\begin{align}
\Pr\Bigg\{\bigcap\limits_{\ell=1}^{L-1}\bigg\{\sum\limits_{i=1}^{n} \tilde f_{\ell,i}(W, \boldsymbol{Y}^{i-1})^2 \in[n(s_\ell-L\gamma), n(s_\ell+\gamma)]\bigg\}\Bigg\}=1.\label{defTildeFeq2*}
\end{align}
On the other hand, it follows from~\eqref{defTildeFeq1**}, \eqref{defTildeFeq2}, the fact $\Pr\left\{\sum_{\ell=1}^L \sum_{k=1}^n f_{\ell,i}(W, \boldsymbol{Y}^{i-1})^2  \le n P\right\}=1$ and the assumption $\sum_{\ell=1}^L s_\ell = P$ that
\begin{align}
\Pr\Bigg\{\hspace{-0.25 in}\sum\limits_{\substack{(\ell,i)\in\\\hspace{0.25 in} \mathcal{L}\times\{1, 2, \ldots, n\} \setminus\{(L, n)\}}} \hspace{-0.55 in}\tilde f_{\ell,i}(W, \boldsymbol{Y}^{i-1})^2  \le nP\Bigg\}=1. \label{defTildeFeq2**}
\end{align}
Finally, in view of~\eqref{defTildeFeq2**}, we define $\tilde f_{L,n}$ as
 \begin{align}
&\tilde f_{L,n}(w, \mathbf{y}^{n-1})\notag\\*
&=\begin{cases}
f_{L,n}(w, \mathbf{y}^{n-1}) & \parbox[t]{4 in}{if $f_{L,n}(w, \mathbf{y}^{n-1})^2+\hspace{-0.25 in}\sum\limits_{\substack{(\ell,i)\in\\\hspace{0.25 in} \mathcal{L}\times\{1, 2, \ldots, n\} \setminus\{(L, n)\}}} \hspace{-0.55 in}\tilde f_{\ell,i}(w, \mathbf{y}^{i-1})^2 =nP$,}\\
 \sqrt{nP - \hspace{-0.25 in}\sum\limits_{\substack{(\ell,i)\in\\\hspace{0.25 in} \mathcal{L}\times\{1, 2, \ldots, n\} \setminus\{(L, n)\}}} \hspace{-0.55 in}\tilde f_{\ell,i}(w, \mathbf{y}^{i-1})^2}& \parbox[t]{4 in}{if $f_{L,n}(w, \mathbf{y}^{n-1})^2+\hspace{-0.25 in}\sum\limits_{\substack{(\ell,i)\in\\\hspace{0.25 in} \mathcal{L}\times\{1, 2, \ldots, n\} \setminus\{(L, n)\}}} \hspace{-0.55 in}\tilde f_{\ell,i}(w, \mathbf{y}^{i-1})^2 < nP$.}
 \end{cases} \label{defTildeFeq3}
\end{align}
Combining~\eqref{defTildeFeq2*}, \eqref{defTildeFeq3} and the assumption that $\sum_{\ell=1}^L s_\ell = P$, we have
\begin{align}
\Pr\Bigg\{\bigcap\limits_{\ell=1}^{L}\bigg\{\sum\limits_{i=1}^{n} \tilde f_{\ell,i}(W, \boldsymbol{Y}^{i-1})^2 \in [n(s_\ell-L\gamma), n(s_\ell+L^2\gamma)]\bigg\}\cap\Bigg\{\sum\limits_{\ell=1}^{L}\sum\limits_{i=1}^{n}\tilde f_{\ell,i}(W, \boldsymbol{Y}^{i-1})^2  = nP\Bigg\}\Bigg\}=1. \label{defTildeFeq3*}
\end{align}
\textbf{3)} A decoding function
\[
\tilde \varphi = \varphi
\]
for providing an estimate of $W$ at node~$\mathrm{d}$. \hfill $\blacksquare$
\end{Definition}
\medskip
\begin{Remark}
The basic idea behind transforming a code in Definition~\ref{definitionTransformedCode} is simple. Suppose we are given an $(n, M, P)$-feedback code, a $\gamma \ge 0$ and an $\mathbf{s}=[s_1\ s_2\ \ldots \ s_L]\in \mathbb{R}_+^L$ such that $\sum_{\ell=1}^L s_\ell = P$. Then, the $(\gamma, \mathbf{s})$-modified code is formed by
\begin{enumerate}
\item[(i)] truncating a transmitted codeword if the power transmitted over the $\ell^{\text{th}}$ channel exceeds $n(s_\ell + \gamma)$, which can be seen from~\eqref{defTildeFeq1} and the third clause of~\eqref{defTildeFeq2};
     \item[(ii)] boosting the power of the transmitted codeword if the power transmitted over the $\ell^{\text{th}}$ channel falls below $n(s_\ell -L \gamma)$, which can be seen from the second clause of~\eqref{defTildeFeq2};
         \item[(iii)] adjusting the last symbol transmitted over the $L^{\text{th}}$ channel (i.e., $X_{L, n}$) so that the total transmitted power is exactly equal to~$nP$, which can be seen from the second clause of~\eqref{defTildeFeq3}.
\end{enumerate}
\end{Remark}
\medskip

Given an $(n, M, P)$-feedback code, we consider the corresponding $(\gamma, \mathbf{s})$-modified code constructed in Definition~\ref{definitionTransformedCode} and let $\tilde p_{W, \boldsymbol{X}^n, \boldsymbol{Y}^n, \hat W}$ be the distribution induced by the modified code. By~\eqref{defTildeFeq3*}, we see that
\begin{align}
\Pr_{\tilde p_{\boldsymbol{X}^n}}\left\{\bigcap\limits_{\ell=1}^{L}\bigg\{\sum\limits_{k=1}^{n} X_{\ell, k}^2 \in[n(s_\ell-L^2\gamma), n(s_\ell+L\gamma)]\bigg\}\cap\bigg\{\sum\limits_{\ell=1}^{L}\sum\limits_{k=1}^{n}X_{\ell, k}^2  = nP\bigg\}\right\}=1. \label{modifiedCodeProperty*}
\end{align}
Define the $\Delta$-bounding box
\begin{equation}
\Gamma^{(\Delta)}(\mathbf{s})\stackrel{\text{def}}{=} [s_1-\Delta,s_1+\Delta]\times [s_2-\Delta,s_2+\Delta]\times \ldots \times [s_L-\Delta,s_L+\Delta]. \label{defGammaS}
\end{equation}
for each $\gamma\ge 0$ and each $\mathbf{s}\in \mathbb{R}_+^L$.
It then follows from~\eqref{modifiedCodeProperty*} that
\begin{align}
\Pr_{\tilde p_{\boldsymbol{X}^n}}\left\{\left\{\phi(\boldsymbol{X}^n)\in\Gamma^{(L^2\gamma)}(\mathbf{s})\right\}\cap\bigg\{\sum\limits_{\ell=1}^{L}\sum\limits_{k=1}^{n}X_{\ell, k}^2  = nP\bigg\}\right\}=1. \label{modifiedCodeProperty}
\end{align}
The following lemma is a natural consequence of Definition~\ref{definitionTransformedCode}, and the proof is deferred to Appendix~\ref{appendixA}.
\medskip
\begin{Lemma}\label{lemmaTransformedCode}
Given an $(n, M, P)$-feedback code, let $p_{\boldsymbol{X}^n, \boldsymbol{Y}^n}$ be the distribution induced by the code. Fix any $\gamma\ge 0$ and any $\mathbf{s}\in \mathbb{R}_+^L$ such that $\sum_{\ell=1}^L s_\ell = P$, and let $\tilde p_{\boldsymbol{X}^n, \boldsymbol{Y}^n}$ be the distribution induced by the $(\gamma, \mathbf{s})$-modified code based on the $(n, M, P)$-feedback code. Then, we have
\begin{align}
\int_{\mathcal{A}}p_{\boldsymbol{X}^n, \boldsymbol{Y}^n}(\mathbf{x}^n, \mathbf{y}^n)\mathbf{1}\left\{\phi(\mathbf{x}^n)\in \Gamma^{(\gamma)}(\mathbf{s})\right\} \mathbf{1}\left\{\sum_{\ell=1}^L\sum_{k=1}^n x_{\ell,k}^2=nP\right\}\mathrm{d}\mathrm{x}^n\mathrm{d}\mathrm{y}^n \le \int_{\mathcal{A}} \tilde p_{\boldsymbol{X}^n, \boldsymbol{Y}^n}(\mathbf{x}^n, \mathbf{y}^n)\mathrm{d}\mathrm{x}^n\mathrm{d}\mathrm{y}^n \label{lemmaTransformedCodeSt}
\end{align}
for all Borel measurable~$\mathcal{A}\subseteq\mathbb{R}^{L\times n}\times \mathbb{R}^{L\times n}$.
\end{Lemma}

\subsection{Curtiss' Theorem} \label{sectionCurtiss}
Curtiss' theorem~\cite[Th.~3]{curtiss} states that convergence of moment generating functions leads to convergence in distribution. The formal statement is reproduced below.
\smallskip
 \begin{Theorem}[Curtiss' theorem] \label{thmCurtiss}
 Let $U^{(n)}$ be a sequence of real-valued random variables. Suppose there exists a random variable $V$ such that
 \begin{align}
\lim_{n\rightarrow\infty} \E\left[e^{tU^{(n)}}\right] = \E\left[e^{t V}\right] \label{st1ThmCurtiss}
 \end{align}
 for all $t\in\mathbb{R}$. Then,
 \begin{align}
\lim_{n\rightarrow\infty} \Pr\{U^{(n)}\le a\} =  \Pr\{V \le a\}
 \end{align}
 for every $a\in\mathbb{R}$ at which $a \mapsto \Pr\{V \le a\}$ is continuous.
 \end{Theorem}
\smallskip

 In contrast to the more well-known L\'evy's continuity theorem~\cite[Sec.~18.1]{Williams1991}, \eqref{st1ThmCurtiss} of Theorem~\ref{thmCurtiss} is required to be true for all real rather than purely imaginary~$t$.

\section{Proof of Theorem~\ref{thmMainResult}} \label{sectionMainResult}
Fix an $\varepsilon\in (0,1)$ and choose an arbitrary sequence of $(\bar n, M_{\mathrm{fb}}^*(\bar n, \varepsilon, P), P, \varepsilon)$-feedback codes. Since
\begin{equation}
C_\varepsilon^{\text{fb}} \ge \mathrm{C}(\mathbf{P}^*)
\end{equation}
by~\eqref{wellKnownFBCapacity}, it suffices to show that
\begin{equation}
\liminf\limits_{n\rightarrow \infty}\frac{1}{\sqrt{n}}\big(\log M_{\mathrm{fb}}^*(n, \varepsilon, P)-n\mathrm{C}(\mathbf{P}^*)\big) \le \sqrt{\mathrm{V}(\mathbf{P}^*)}\, \Phi^{-1}(\varepsilon+ \tau)  \label{goalInProofMainResult}
\end{equation}
for all $\tau>0$. To this end, fix an arbitrary $\tau>0$.
\subsection{Discretizing the Power Allocation Vectors by Appending Symbols} \label{stepAinProof}
 Using Definition~\ref{defCode}, we have
\begin{equation*}
\Pr\left\{\sum_{\ell=1}^L\sum_{k=1}^{\bar n}X_{\ell,k}^2 \le \bar nP\right\} = 1 
\end{equation*}
for the chosen $(\bar n, M_{\mathrm{fb}}^*(\bar n, \varepsilon, P), P, \varepsilon)$-feedback code for each~$\bar n\in \mathbb{N}$.
  Given the chosen $(\bar n, M_{\mathrm{fb}}^*(\bar n, \varepsilon, P), P, \varepsilon)$-feedback code, we can always construct an $(\bar n+L, M_{\mathrm{fb}}^*(\bar n, \varepsilon, P), P, \varepsilon)$-feedback code by appending a carefully chosen tuple $(\boldsymbol{X}_{\bar n+1}, \boldsymbol{X}_{\bar n+2}, \ldots, \boldsymbol{X}_{\bar n+L})$ to each transmitted codeword $\boldsymbol{X}^{\bar n}$ generated by the $(\bar n, M_{\mathrm{fb}}^*(\bar n, \varepsilon, P), P, \varepsilon)$-feedback code such that
\begin{equation}
\Pr\left\{\sum_{k=1}^{\bar n+L}X_{\ell,k}^2 = P\left\lceil\frac{1}{P}\sum_{k=1}^{\bar n}X_{\ell,k}^2\right\rceil  \text{ for all $\ell\in \mathcal{L}$}\right\} = 1, 
\end{equation}
which implies that
\begin{equation}
\Pr\left\{\sum_{k=1}^{\bar n+L}X_{\ell,k}^2 \text{ is a multiple of~$P$ for all $\ell\in \mathcal{L}$ and }\sum_{\ell=1}^L\sum_{k=1}^{\bar n+L}X_{\ell,k}^2 \le (\bar n+L)P\right\} = 1. \label{powerConstraintInProof**}
\end{equation}
In addition, given the $(\bar n+L, M_{\mathrm{fb}}^*(\bar n, \varepsilon, P), P, \varepsilon)$-feedback code, we can always construct an $(\bar n+L+1, M_{\mathrm{fb}}^*(\bar n, \varepsilon, P), \linebreak P, \varepsilon)$-feedback code by appending a carefully chosen $\boldsymbol{X}_{\bar n+L+1}$ to each transmitted codeword $\boldsymbol{X}^{\bar n+L}$ generated by the $(\bar n +L, M_{\mathrm{fb}}^*(\bar n, \varepsilon, P), P, \varepsilon)$-feedback code such that
\begin{equation}
\Pr\left\{\sum_{k=1}^{\bar n+L+1}X_{\ell,k}^2 \text{ is a multiple of~$P$ for all $\ell\in \mathcal{L}$ and }\sum_{\ell=1}^L\sum_{k=1}^{\bar n+L+1}X_{\ell,k}^2 = (\bar n+L+1)P\right\} = 1. \label{powerConstraintInProof***}
\end{equation}
To simplify notation, we let
\begin{equation}
n\stackrel{\text{def}}{=}\bar n+L+1. \label{defBarN}
\end{equation}
Construct the set of power allocation vectors
 \begin{align}
 \mathcal{S}^{(n)} \stackrel{\text{def}}{=} \left\{\frac{P}{n}\cdot a^L  \left|  a^L\in \mathbb{Z}_+^L, \:\sum_{\ell=1}^L a_\ell = n \right.\right\}, \label{defSetS}
 \end{align}
 which can be viewed as a set of quantized power allocation vectors $\mathbf{s}$ with quantization level $P/n$ that satisfy the equality power constraint
 \begin{equation*}
 \sum_{\ell=1}^L s_\ell = P.
 \end{equation*}
It follows from~\eqref{defSetS}, \eqref{powerConstraintInProof***} and Definition~\ref{definitionPowerType} that
\begin{equation}
|\mathcal{S}^{(n)}| \le n^L \label{cardinalityBound}
\end{equation}
and
\begin{align}
\Pr\left\{\phi(\boldsymbol{X}^n)\in  \mathcal{S}^{(n)}\right\} = 1. \label{powerConstraintInProof****}
\end{align}

\subsection{Obtaining a Lower Bound on the Error Probability in Terms of the Type-II Error of a Hypothesis Test}
 Let $p_{W,\boldsymbol{X}^n, \boldsymbol{Y}^n, \hat W}$ be the probability distribution induced by the $(n, M_{\mathrm{fb}}^*(\bar n, \varepsilon, P), P, \varepsilon)$-feedback code constructed above for each $n\in\{L+2, L+3, \ldots\}$, where $p_{W, \boldsymbol{X}^n, \boldsymbol{Y}^n, \hat W}$ is obtained according to \eqref{memorylessStatement}.
  Fix an $n\in\{L+2, L+3, \ldots\} $ and the corresponding $(n, M_{\mathrm{fb}}^*(\bar n, \varepsilon, P), P, \varepsilon)$-feedback code.
Recall the definition of $P_\ell$ for each $\ell\in\mathcal{L}$ in~\eqref{pEllValue} and define the distribution
\begin{align}
r_{\boldsymbol{Y}^n, \hat W}\stackrel{\text{def}}{=} p_{\hat W|\boldsymbol{Y}^n} r_{\boldsymbol{Y}^n} \label{defDistS}
\end{align}
where\footnote{We note that even if we exclude the set of power types in the set $\Pi^{(n)}$ which is defined later in~\eqref{defSetPiN}, this leads to another valid definition of $r_{\boldsymbol{Y}^n}(\mathbf{y}^n)$.}
\begin{align}
  r_{\boldsymbol{Y}^n}(\mathbf{y}^n)  \stackrel{\text{def}}{=} \frac{1}{2|\mathcal{S}^{(n)}|}\sum_{\mathbf{s}\in \mathcal{S}^{(n)}}\prod_{\ell=1}^L \prod_{k=1}^n \mathcal{N}(y_{\ell,k}; 0, s_\ell+N_\ell) + \frac{1}{2}\prod_{\ell=1}^L\prod_{k=1}^n \mathcal{N}(y_{\ell,k}; 0, P_\ell+N_\ell). \label{defDistSyk}
  \end{align}
  The choice of $r_{\boldsymbol{Y}^n}$ in~\eqref{defDistSyk} is motivated by the choice of the auxiliary output distribution in~\cite[Sec.~X-A]{Hayashi09} where DMCs are considered.
Then, it follows from Proposition~\ref{propositionBHTLowerBound} and Definition~\ref{defCode} with the identifications $U\equiv W$, $V\equiv \hat W$, $p_{U,V}\equiv p_{W,\hat W}$, $q_V\equiv r_{\hat W}$, $|\mathcal{W}|\equiv M_{\mathrm{fb}}^*(\bar n, \varepsilon, P)$ and $\alpha\equiv \Pr\{\hat W \ne W\} \le \varepsilon$ that
 \begin{align}
\beta_{1-\varepsilon}(p_{W,\hat W}\|p_W r_{\hat W}) \le \beta_{1-\alpha}(p_{W,\hat W}\|p_W r_{\hat W}) \le 1/M_{\mathrm{fb}}^*(\bar n, \varepsilon, P). \label{eqnBHTReverseChain}
 \end{align}
  \subsection{Obtaining a Non-Asymptotic Bound from Simplifying the Type-II Error of the Binary Hypothesis Test}
Using the DPI of $\beta_{1-\varepsilon}$ by introducing $\boldsymbol{X}^n$ and $\boldsymbol{Y}^n$, we have
\begin{align}
\beta_{1-\varepsilon}(p_{W,\hat W}\|p_W r_{\hat W}) \ge \beta_{1-\varepsilon}\bigg(p_{W,\boldsymbol{X}^n,\boldsymbol{Y}^n,\hat W}\bigg\|p_W r_{\boldsymbol{Y}^n,\hat W}\prod_{k=1}^n p_{\boldsymbol{X}_k|W,\boldsymbol{Y}^{k-1}}\bigg) \label{eqnBHTFirstChain}
\end{align}
%
where
 \begin{align}
   p_{W, \boldsymbol{X}^n , \boldsymbol{Y}^n, \hat W} =p_W p_{\hat W|\boldsymbol{Y}^n}\prod_{k=1}^n (p_{\boldsymbol{X}_k|W, \boldsymbol{Y}^{k-1}}p_{\boldsymbol{Y}_k|\boldsymbol{X}_k}) \label{eqnBHTFirstChain*}
 \end{align}
by~\eqref{memorylessStatement}.
%
 Combining \eqref{eqnBHTFirstChain}, \eqref{eqnBHTFirstChain*} and~\eqref{defDistS}, we have
 \begin{align}
&  \beta_{1-\varepsilon}(p_{W,\hat W}\|p_W r_{\hat W}) \notag\\
& \ge  \beta_{1-\varepsilon}\left(p_W p_{\hat W|Y^n}\prod_{k=1}^n (p_{\boldsymbol{X}_k|\boldsymbol{Y}^{k-1}, W}p_{\boldsymbol{Y}_k|\boldsymbol{X}_k})\left\|p_W r_{\boldsymbol{Y}^n} p_{\hat W|Y^n} \prod_{k=1}^n p_{\boldsymbol{X}_k|W,\boldsymbol{Y}^{k-1}} \right.\right). \label{eqnBHT5thChain}
 \end{align}
Fix any constant $\xi_n>0$ to be specified later. Using Lemma~\ref{lemmaDPI}, \eqref{eqnBHT5thChain} and~\eqref{defChannelInDefinition*},
we have
\begin{align}
 \beta_{1-\varepsilon}(p_{W,\hat W}\|p_W r_{\hat W})  \ge \frac{1}{\xi_n}\left(1-\varepsilon - \Pr_{p_{\boldsymbol{X}^n, \boldsymbol{Y}^n}}\left\{\frac{\prod_{k=1}^n q_{\boldsymbol{Y}|\boldsymbol{X}}(\boldsymbol{Y}_k|\boldsymbol{X}_{k})}{r_{\boldsymbol{Y}^n}(\boldsymbol{Y}^n)} \ge \xi_n \right\}\right), \label{eqnBHTThirdChain}
\end{align}
which together with \eqref{eqnBHTReverseChain} implies that
\begin{align}
\log M_{\mathrm{fb}}^*(\bar n, \varepsilon, P)
 \le \log\xi_n - \log\left(1-\varepsilon -\Pr_{p_{\boldsymbol{X}^n, \boldsymbol{Y}^n}}\left\{\log\bigg(\frac{\prod_{k=1}^n q_{\boldsymbol{Y}|\boldsymbol{X}}(\boldsymbol{Y}_k|\boldsymbol{X}_{k})}{r_{\boldsymbol{Y}^n}(\boldsymbol{Y}^n)}\bigg) \ge \log \xi_n \right\}\right).\label{eqnBHTSecondChain}
\end{align}
\subsection{Splitting the Probability Term into Multiple Terms Corresponding to Different Power Types of $\boldsymbol{X}^n$}
Define\footnote{The conclusion of this proof remains unchanged if the $n^{1/6}$ term in~\eqref{defSetPiN} is replaced by $n^a$ for any $a\in(0, 1/2)$.}
\begin{align}
\Pi^{(n)} \stackrel{\text{def}}{=} \left\{\mathbf{s}\in\mathcal{S}^{(n)}  \left|  \|\mathbf{s}-\mathbf{P}^*\|_2\le \frac{1}{n^{1/6}} \right.\right\} \label{defSetPiN}
\end{align}
to be the set of power allocation vectors in $\mathcal{S}^{(n)}$ that are close to the optimal power allocation vector $\mathbf{P}^*$ (cf.\ \eqref{pEllValue*}).
Following~\eqref{eqnBHTSecondChain}, we use~\eqref{powerConstraintInProof****} to obtain
\begin{align}
&\Pr_{p_{\boldsymbol{X}^n, \boldsymbol{Y}^n}}\left\{\log\bigg(\frac{\prod_{k=1}^n q_{\boldsymbol{Y}|\boldsymbol{X}}(\boldsymbol{Y}_k|\boldsymbol{X}_{k})}{r_{\boldsymbol{Y}^n}(\boldsymbol{Y}^n)}\bigg) \ge \log \xi_n \right\}\notag\\
& = \Pr_{p_{\boldsymbol{X}^n, \boldsymbol{Y}^n}}\left\{\left\{\log\bigg(\frac{\prod_{k=1}^n q_{\boldsymbol{Y}|\boldsymbol{X}}(\boldsymbol{Y}_k|\boldsymbol{X}_{k})}{r_{\boldsymbol{Y}^n}(\boldsymbol{Y}^n)}\bigg) \ge \log \xi_n\right\}\cap\left\{\phi(\boldsymbol{X}^n)\in \Pi^{(n)} \right\} \right\} \notag\\
&\qquad + \sum_{\mathbf{s}\in \mathcal{S}^{(n)}\setminus\Pi^{(n)}}\Pr_{p_{\boldsymbol{X}^n, \boldsymbol{Y}^n}}\left\{\left\{\log\bigg(\frac{\prod_{k=1}^n q_{\boldsymbol{Y}|\boldsymbol{X}}(\boldsymbol{Y}_k|\boldsymbol{X}_{k})}{r_{\boldsymbol{Y}^n}(\boldsymbol{Y}^n)}\bigg) \ge \log \xi_n\right\}\cap\left\{\phi(\boldsymbol{X}^n)=\mathbf{s} \right\} \right\}. \label{eqnBHT3rdChain}
\end{align}
In order to bound the first term in~\eqref{eqnBHT3rdChain}, we let
\begin{equation}
\gamma\stackrel{\text{def}}{=}\frac{1}{n^{1/6}} \label{defgammaInProof}
 \end{equation}
 and define $p_{\boldsymbol{X}^n, \boldsymbol{Y}^n}^{*}$ be the distribution induced by the $(\gamma, \mathbf{P}^*)$-modified code based on the $(n, M_{\mathrm{fb}}^*(\bar n, \varepsilon, P), P, \varepsilon)$-feedback code defined in Definition~\ref{definitionTransformedCode}.
 Then, consider the following chain of inequalities:
\begin{align}
&\Pr_{p_{\boldsymbol{X}^n, \boldsymbol{Y}^n}}\left\{\left\{\log\bigg(\frac{\prod_{k=1}^n q_{\boldsymbol{Y}|\boldsymbol{X}}(\boldsymbol{Y}_k|\boldsymbol{X}_{k})}{r_{\boldsymbol{Y}^n}(\boldsymbol{Y}^n)}\bigg) \ge \log \xi_n\right\}\cap\left\{\phi(\boldsymbol{X}^n)\in \Pi^{(n)} \right\}\right\}\notag\\*
&\le \Pr_{p_{\boldsymbol{X}^n, \boldsymbol{Y}^n}^{*}}\left\{\log\bigg(\frac{\prod_{k=1}^n q_{\boldsymbol{Y}|\boldsymbol{X}}(\boldsymbol{Y}_k|\boldsymbol{X}_{k})}{r_{\boldsymbol{Y}^n}(\boldsymbol{Y}^n)}\bigg) \ge \log \xi_n\right\} \label{eqnBHT3rd1Chain(a)} \\*
& \le \Pr_{p_{\boldsymbol{X}^n, \boldsymbol{Y}^n}^{*}}\left\{\sum_{k=1}^n\log\bigg(\frac{ q_{\boldsymbol{Y}|\boldsymbol{X}}(\boldsymbol{Y}_k|\boldsymbol{X}_{k})}{\prod_{\ell=1}^L\mathcal{N}(Y_{\ell,k};0, P_\ell+N_\ell)}\bigg) \ge \log \xi_n - \log 2\right\} \label{eqnBHT3rd1Chain(b)}
\end{align}
where
\begin{itemize}
\item \eqref{eqnBHT3rd1Chain(a)} is due to Lemma~\ref{lemmaTransformedCode} and the fact that $\Pr_{p_{\boldsymbol{X}^n, \boldsymbol{Y}^n}}\left\{\sum_{\ell=1}^L\sum_{k=1}^n X_{\ell,k}^2=nP\right\}=1$ (cf.\ \eqref{defSetS} and~\eqref{powerConstraintInProof****}).
\item \eqref{eqnBHT3rd1Chain(b)} is due to the definition of $r_{\boldsymbol{Y}^n}$ in~\eqref{defDistSyk}.
\end{itemize}
Similarly, in order to bound the second term in~\eqref{eqnBHT3rdChain}, we let $p_{\boldsymbol{X}^n, \boldsymbol{Y}^n}^{(\mathbf{s})}$ be the distribution induced by the $(0, \mathbf{s})$-modified code and consider the following chain of inequalities for each $\mathbf{s}\in \mathcal{S}^{(n)}\setminus\Pi^{(n)}$:
\begin{align}
&\Pr_{p_{\boldsymbol{X}^n, \boldsymbol{Y}^n}}\left\{\left\{\log\bigg(\frac{\prod_{k=1}^n q_{\boldsymbol{Y}|\boldsymbol{X}}(\boldsymbol{Y}_k|\boldsymbol{X}_{k})}{r_{\boldsymbol{Y}^n}(\boldsymbol{Y}^n)}\bigg) \ge \log \xi_n\right\}\cap\left\{\phi(\boldsymbol{X}^n)=\mathbf{s} \right\}\right\}\notag\\
&\le \Pr_{p_{\boldsymbol{X}^n, \boldsymbol{Y}^n}^{(\mathbf{s})}}\left\{\log\bigg(\frac{\prod_{k=1}^n q_{\boldsymbol{Y}|\boldsymbol{X}}(\boldsymbol{Y}_k|\boldsymbol{X}_{k})}{r_{\boldsymbol{Y}^n}(\boldsymbol{Y}^n)}\bigg) \ge \log \xi_n\right\} \label{eqnBHT4thChain(a)} \\
& \le \Pr_{p_{\boldsymbol{X}^n, \boldsymbol{Y}^n}^{(\mathbf{s})}}\left\{\sum_{k=1}^n\log\bigg(\frac{ q_{\boldsymbol{Y}|\boldsymbol{X}}(\boldsymbol{Y}_k|\boldsymbol{X}_{k})}{\prod_{\ell=1}^L\mathcal{N}(Y_{\ell,k};0, s_\ell+N_\ell)}\bigg) \ge \log \xi_n - \log\big(2|\mathcal{S}^{(n)}|\big)\right\} \label{eqnBHT4thChain(b)}
\end{align}
where
\begin{itemize}
\item \eqref{eqnBHT4thChain(a)} is due to Lemma~\ref{lemmaTransformedCode}.
\item \eqref{eqnBHT4thChain(b)} is due to the definition of $r_{\boldsymbol{Y}^n}$ in~\eqref{defDistSyk}.
\end{itemize}
Combining~\eqref{eqnBHT3rdChain}, \eqref{eqnBHT3rd1Chain(b)}, \eqref{eqnBHT4thChain(b)} and the definition of $q_{\boldsymbol{Y}|\boldsymbol{X}}$ in~\eqref{defAWGNchannelParallel} followed by letting
\begin{equation}
Z_{\ell,k}\stackrel{\text{def}}{=}Y_{\ell,k}-X_{\ell,k}
 \end{equation}
 for each $\ell\in\mathcal{L}$ and each $k\in\{1, 2, \ldots, n\}$, we obtain
\begin{align}
&\Pr_{p_{\boldsymbol{X}^n, \boldsymbol{Y}^n}}\left\{\log\bigg(\frac{\prod_{k=1}^n q_{\boldsymbol{Y}|\boldsymbol{X}}(\boldsymbol{Y}_k|\boldsymbol{X}_{k})}{r_{\boldsymbol{Y}^n}(\boldsymbol{Y}^n)}\bigg) \ge \log \xi_n \right\}\notag\\*
& \le \Pr_{p_{\boldsymbol{X}^n, \boldsymbol{Y}^n}^{*}}\left\{n\mathrm{C}(\mathbf{P}^*)+ \sum_{k=1}^n\sum_{\ell=1}^L\frac{-\big(\frac{P_\ell}{N_\ell} \big) Z_{\ell,k}^2 + 2X_{\ell,k}Z_{\ell,k}+X_{\ell,k}^2}{2(P_\ell+N_\ell)} \ge \log \xi_n - \log 2\right\} \notag\\*
&\qquad + \sum_{\mathbf{s}\in \mathcal{S}^{(n)}\setminus\Pi^{(n)}}\Pr_{p_{\boldsymbol{X}^n, \boldsymbol{Y}^n}^{(\mathbf{s})}}\left\{n\mathrm{C}(\mathbf{s})+\sum_{k=1}^n\sum_{\ell=1}^L\frac{-\big(\frac{s_\ell}{N_\ell}\big) Z_{\ell,k}^2 + 2X_{\ell,k}Z_{\ell,k}+X_{\ell,k}^2}{2(s_\ell+N_\ell)} \ge \log \xi_n - \log\big(2|\mathcal{S}^{(n)}|\big)\right\} \label{eqnBHT5th1Chain}
\end{align}
where $\mathrm{C}(\cdot)$ is as defined in~\eqref{defCPparallelNFB}.
In order to simplify the RHS of~\eqref{eqnBHT5th1Chain}, we define $\xi_n>0$ such that
\begin{align}
\log \xi_n \stackrel{\text{def}}{=} n\mathrm{C}(\mathbf{P}^*) + \sqrt{n}\left(\sqrt{\mathrm{V}(\mathbf{P}^*)}\, \Phi^{-1}(\varepsilon+\tau)\right) + \log\big(2|\mathcal{S}^{(n)}|\big). \label{defxin}
\end{align}
In addition, for each $\mathbf{d}\in \mathbb{R}_+^L$, let
\begin{align}
U_k^{(\mathbf{d})} \stackrel{\text{def}}{=}  \sum_{\ell=1}^L\frac{-\big(\frac{d_\ell}{N_\ell}\big) Z_{\ell,k}^2 + 2X_{\ell,k}Z_{\ell,k}+d_\ell}{2(d_\ell+N_\ell)} \label{defUkd}
\end{align}
for each $k\in\{1, 2, \ldots, n\}$.
By using~\eqref{eqnBHT5th1Chain}, \eqref{defxin} and \eqref{defUkd} together with the facts by~\eqref{modifiedCodeProperty} that 
\begin{align}
\Pr_{ p_{\boldsymbol{X}^n,\boldsymbol{Y}^n}^*}\left\{\left\{\phi(\boldsymbol{X}^n)\in\Gamma^{(L^2\gamma)}(\mathbf{P}^*)\right\}\cap\bigg\{\sum\limits_{\ell=1}^{L}\sum\limits_{k=1}^{n}X_{\ell, k}^2  = nP\bigg\}\right\}=1 \label{defDistP*}
\end{align}
and
\begin{align}
\Pr_{ p_{\boldsymbol{X}^n,\boldsymbol{Y}^n}^{(\mathbf{s})}}\left\{\phi(\boldsymbol{X}^n)\in\Gamma^{(0)}(\mathbf{s})\right\}=1 \label{defDistPs}
\end{align}
for each $\mathbf{s}\in \mathcal{S}^{(n)}$,
we can express \eqref{eqnBHT5th1Chain} as
\begin{align}
&\Pr_{p_{\boldsymbol{X}^n, \boldsymbol{Y}^n}}\left\{\log\bigg(\frac{\prod_{k=1}^n q_{\boldsymbol{Y}|\boldsymbol{X}}(\boldsymbol{Y}_k|\boldsymbol{X}_{k})}{r_{\boldsymbol{Y}^n}(\boldsymbol{Y}^n)}\bigg) \ge \log \xi_n \right\}\notag\\*
&\quad\le \Pr_{p_{\boldsymbol{X}^n, \boldsymbol{Y}^n}^{*}}\left\{\frac{1}{\sqrt{n \mathrm{V}(\mathbf{P}^*)}}\sum_{k=1}^n U_k^{(\mathbf{P}^*)}  \ge \Phi^{-1}(\varepsilon+\tau)\right\} \notag\\
&\qquad + \sum_{\mathbf{s}\in \mathcal{S}^{(n)}\setminus\Pi^{(n)}}\Pr_{p_{\boldsymbol{X}^n, \boldsymbol{Y}^n}^{(\mathbf{s})}}\left\{\frac{1}{\sqrt{n}}\sum_{k=1}^n U_k^{(\mathbf{s})}\ge \sqrt{n}\big(\mathrm{C}(\mathbf{P}^*)-\mathrm{C}(\mathbf{s})\big)+ \sqrt{\mathrm{V}(\mathbf{P}^*)}\, \Phi^{-1}(\varepsilon+\tau)\right\}. \label{eqnBHT5th2Chain}
\end{align}
\subsection{Applying Curtiss' Theorem When $\phi(\boldsymbol{X}^n)$ is Close to $\mathbf{P}^*$} \label{stepBInProof}
In order to simplify the first term in~\eqref{eqnBHT5th2Chain}, we
define
\begin{align}
V_k^{(\mathbf{P}^*)} \stackrel{\text{def}}{=}  \sum_{\ell=1}^L\frac{-\big(\frac{P_\ell}{N_\ell}\big) Z_{\ell,k}^2 + 2\sqrt{P_\ell}Z_{\ell,k}+P_\ell}{2(P_\ell+N_\ell)} \label{defVkd}
\end{align}
for each $k\in\{1, 2, \ldots, n\}$
and want to show that
\begin{align}
\lim_{n\rightarrow \infty}\E_{p_{\boldsymbol{X}^n, \boldsymbol{Y}^n}^{*}}\left[e^{\frac{t}{\sqrt{n }}\sum\limits_{k=1}^n U_k^{(\mathbf{P}^*)}}\right] = \lim_{n\rightarrow \infty}\E_{p_{\boldsymbol{Z}^n}^{*}}\left[e^{\frac{t}{\sqrt{n }}\sum\limits_{k=1}^n V_k^{(\mathbf{P}^*)}}\right] \label{levyThmEq1}
\end{align}
for all $t\in \mathbb{R}$ where
\begin{align}
p_{\boldsymbol{Z}^n}^{*}(\mathrm{z}^n) \triangleq \prod_{k=1}^n \mathcal{N} (z_{\ell,k}; 0, N_\ell).
\end{align}
 To this end, recall the following statements due to the channel law: 
 \begin{enumerate}
 \item[(i)] $Z_{\ell,k}\sim\mathcal{N}(z_{\ell,k};0, N_\ell)$ for all $\ell\in \mathcal{L}$ and all $k\in\{1, 2, \ldots, n\}$;
 \item[(ii)] $\{Z_{\ell,k}|\ell\in\mathcal{L}, k\in\{1, 2, \ldots, n\}\}$ are independent;
 \item[(iii)] $\boldsymbol{Z}_k$ and $(\boldsymbol{X}^{k},\boldsymbol{Y}^{k-1},\boldsymbol{Z}^{k-1})$ are independent for all $k\in\{1, 2, \ldots, n\}$.
 \end{enumerate}
For any $t\in\mathbb{R}$ and any~$n\in\{L+2, L+3, \ldots\}$ such that $n\ge t^2$, since
\begin{align}
P_\ell - L^2\gamma-\frac{1}{n}\sum\limits_{k=1}^n X_{\ell,k}^2 \le 0 \le P_\ell + L^2\gamma-\frac{1}{n}\sum\limits_{k=1}^n X_{\ell,k}^2
\end{align}
by~\eqref{defDistP*} and $P_\ell+N_\ell+\frac{tP_\ell}{\sqrt{n}} >0$
for all $\ell\in\mathcal{L}$, we have
 \begin{align}
& \E_{p_{\boldsymbol{X}^n, \boldsymbol{Y}^n}^{*}}\Biggl[e^{\frac{t}{\sqrt{n}}\sum\limits_{k=1}^n U_k^{(\mathbf{P}^*)}}\cdot e^{t^2\sum\limits_{\ell=1}^L\frac{N_\ell \left(P_\ell - L^2\gamma-\frac{1}{n}\sum\limits_{k=1}^n X_{\ell,k}^2\right) }{2(P_\ell+N_\ell)\left(P_\ell+N_\ell+\frac{tP_\ell}{\sqrt{n }}\right)}}\:\Biggr] \notag\\*
 &\le \E_{p_{\boldsymbol{X}^n, \boldsymbol{Y}^n}^{*}}\left[e^{\frac{t}{\sqrt{n }}\sum\limits_{k=1}^n U_k^{(\mathbf{P}^*)}}\right] \\*
  &\le \E_{p_{\boldsymbol{X}^n, \boldsymbol{Y}^n}^{*}}\Biggl[e^{\frac{t}{\sqrt{n}}\sum\limits_{k=1}^n U_k^{(\mathbf{P}^*)}}\cdot e^{t^2\sum\limits_{\ell=1}^L\frac{N_\ell \left(P_\ell + L^2\gamma-\frac{1}{n}\sum\limits_{k=1}^n X_{\ell,k}^2\right) }{2(P_\ell+N_\ell)\left(P_\ell+N_\ell+\frac{tP_\ell}{\sqrt{n }}\right)}}\:\Biggr]. \label{levyThmEq1*}
 \end{align}
In order to simplify the above chain of inequalities, we need the following lemma, whose proof is deferred to Appendix~\ref{appendixC} because it involves straightforward calculations based on~\eqref{defUkd}, \eqref{defVkd} and the channel law.
 \smallskip
 \begin{Lemma} \label{lemma1InProof}
 For any $\lambda\in\mathbb{R}$, we have
 \begin{align}
 \E_{p_{\boldsymbol{X}^n, \boldsymbol{Y}^n}^{*}}\Biggl[e^{\lambda\sum\limits_{k=1}^n U_k^{(\mathbf{P}^*)}}\cdot e^{\lambda^2\sum\limits_{\ell=1}^L\frac{N_\ell \left(nP_\ell  -\sum\limits_{k=1}^n X_{\ell,k}^2\right) }{2(P_\ell+N_\ell)\left((1+\lambda)P_\ell+N_\ell\right)}}\:\Biggr] = \E_{p_{\boldsymbol{Z}^n}^{*}}\left[e^{\lambda\sum\limits_{k=1}^n V_k^{(\mathbf{P}^*)}}\right]. \label{appendixCeq1}
 \end{align}
 \end{Lemma}

Lemma~\ref{lemma1InProof}, which forms the crux of the proof of Theorem \ref{thmMainResult}, is important because it
 establishes the equivalence in distribution between the sum $\sum_{k=1}^n U_k^{(\mathbf{P}^*)}$, which contains {\em dependent} random variables, and the sum $\sum_{k=1}^n V_k^{(\mathbf{P}^*)}$, which contains {\em independent} random variables. The former is intractable to analyze while the latter can be analyzed in a straightforward manner by invoking the central limit theorem.
 %

Using Lemma~\ref{lemma1InProof}, we can simplify~\eqref{levyThmEq1*} through the identification $\lambda\equiv \frac{t}{\sqrt{n}}$ and obtain
\begin{align}
&\E_{p_{\boldsymbol{Z}^n}^{*}}\left[e^{\frac{t}{\sqrt{n }}\sum\limits_{k=1}^n V_k^{(\mathbf{P}^*)}}\right] \cdot e^{-t^2 L^2\gamma \sum\limits_{\ell=1}^L\frac{N_\ell }{2(P_\ell+N_\ell)\left(P_\ell+N_\ell+\frac{tP_\ell}{\sqrt{n }}\right)}}  \notag\\*
&\le \E_{p_{\boldsymbol{X}^n, \boldsymbol{Y}^n}^{*}}\left[e^{\frac{t}{\sqrt{n }}\sum\limits_{k=1}^n U_k^{(\mathbf{P}^*)}}\right]
\\*
&\le \E_{p_{\boldsymbol{Z}^n}^{*}}\left[e^{\frac{t}{\sqrt{n }}\sum\limits_{k=1}^n V_k^{(\mathbf{P}^*)}}\right] \cdot e^{t^2 L^2\gamma \sum\limits_{\ell=1}^L\frac{  N_\ell }{2(P_\ell+N_\ell)\left(P_\ell+N_\ell+\frac{tP_\ell}{\sqrt{n }}\right)}}.
\label{levyThmEq2}
\end{align}
Combining \eqref{levyThmEq2} and~\eqref{defgammaInProof}, we conclude that \eqref{levyThmEq1} holds for each $t\in\mathbb{R}$. Since the moment generating functions of $\frac{1}{\sqrt{n }}\sum_{k=1}^n V_k^{(\mathbf{P}^*)}$ and $\frac{1}{\sqrt{n }}\sum_{k=1}^n U_k^{(\mathbf{P}^*)}$ converge to the same function, it follows from Curtiss' theorem~\cite[Th.~3]{curtiss} (as stated in Theorem~\ref{thmCurtiss}) that
\begin{align}
\lim_{n\rightarrow \infty} \Pr_{p_{\boldsymbol{X}^n, \boldsymbol{Y}^n}^{*}}\left\{\frac{1}{\sqrt{n \mathrm{V}(\mathbf{P}^*)}}\sum_{k=1}^n U_k^{(\mathbf{P}^*)}  \ge \Phi^{-1}(\varepsilon+\tau)\right\} = \lim_{n\rightarrow \infty} \Pr_{p_{\boldsymbol{Z}^n}^{*}}\left\{\frac{1}{\sqrt{n \mathrm{V}(\mathbf{P}^*)}}\sum_{k=1}^n V_k^{(\mathbf{P}^*)}  \ge \Phi^{-1}(\varepsilon+\tau)\right\}. \label{levyThmEq3}
\end{align}
Recognizing that $\big\{V_k^{(\mathbf{P}^*)} \big\}_{k=1}^\infty$ are independent zero-mean Gaussian random variables with variance $\mathrm{V}(\mathbf{P}^*)$ by the definition of $V_k^{(\mathbf{P}^*)} $ in~\eqref{defVkd} and the definition of $\mathrm{V}(\mathbf{P}^*)$ in~\eqref{defVP}, we apply the central limit theorem~\cite{feller} and obtain
\begin{align}
\lim_{n\rightarrow \infty} \Pr_{p_{\boldsymbol{Z}^n}^{*}}\left\{\frac{1}{\sqrt{n \mathrm{V}(\mathbf{P}^*)}}\sum_{k=1}^n V_k^{(\mathbf{P}^*)} \le \Phi^{-1}(\varepsilon+\tau)\right\} = \varepsilon+\tau,
\end{align}
which together with~\eqref{levyThmEq3} implies that
\begin{align}
\lim_{n\rightarrow \infty} \Pr_{p_{\boldsymbol{X}^n, \boldsymbol{Y}^n}^{*}}\left\{\frac{1}{\sqrt{n \mathrm{V}(\mathbf{P}^*)}}\sum_{k=1}^n U_k^{(\mathbf{P}^*)}  \ge \Phi^{-1}(\varepsilon+\tau)\right\} = 1-\varepsilon - \tau. \label{levyThmEq4}
\end{align}
\subsection{Applying Large Deviation Bounds When $\phi(\boldsymbol{X}^n)$ is Far from $\mathbf{P}^*$} \label{stepCInProof}
In order to bound the second term in~\eqref{eqnBHT5th2Chain}, we consider a fixed~$n\in\{L+2, L+3, \ldots\}$ and want to show that there exists some $\kappa>0$ such that
\begin{align}
\mathrm{C}(\mathbf{P}^*)-\mathrm{C}(\mathbf{s}) \ge \kappa\|\mathbf{P}^*-\mathbf{s}\|_{2}^2 \label{eqnBHT6thChain}
\end{align}
for all $\mathbf{s}\in \mathcal{S}^{(n)}$. To this end, we first define the Lagrangian function $f: \mathbb{R}^L \rightarrow \mathbb{R}$ as
\begin{align}
f(\mathbf{d}) \stackrel{\text{def}}{=}   \mathrm{C}(\mathbf{d}) - \frac{1}{2\Lambda}\left(\sum_{\ell=1}^L d_\ell - P\right) + \sum_{\ell=1}^L \mu_\ell d_\ell \label{defFunctionF}
\end{align}
where $\Lambda\ge 0$ is the unique number that satisfies~\eqref{sumPell=P} and~\eqref{pEllValue} and $\mu_\ell\ge 0$ is defined for each $\ell\in\mathcal{L}$ as
\begin{align}
\mu_\ell \stackrel{\text{def}}{=}
\begin{cases}
0 & \text{if $P_\ell>0$,}\\
\frac{1}{2\Lambda}-\frac{1}{2N_\ell} & \text{if $P_\ell=0$.}
\end{cases} \label{defMu}
\end{align}
 Define $N_\text{max} \stackrel{\text{def}}{=} \max\limits_{\ell\in\mathcal{L}}N_\ell$. Then for all $\mathbf{s}\in \mathcal{S}^{(n)}$, we use Taylor's theorem to obtain
\begin{align}
 f(\mathbf{s}) = f(\mathbf{P}^*) + (\mathbf{s}-\mathbf{P}^*)^t\triangledown f(\mathbf{P}^*) +  \frac{1}{2}(\mathbf{s}-\mathbf{P}^*)^t\triangledown^2 f(\bar{\mathbf{s}}) (\mathbf{s}-\mathbf{P}^*) \label{eqnTaylor}
\end{align}
for some $\bar{\mathbf{s}}$ that lies on the line that connects $\mathbf{s}$ and $\mathbf{P}^*$, where $\triangledown f(\mathbf{P}^*)$ denotes the gradient which satisfies
\begin{align}
\triangledown f(\mathbf{P}^*) = 0 \label{eqnTaylor1st}
\end{align}
and $\triangledown^2 f(\bar{\mathbf{s}})$ denotes the Hessian matrix that satisfies
\begin{align}
(\mathbf{s}-\mathbf{P}^*)^t\triangledown^2 f(\bar{\mathbf{s}}) (\mathbf{s}-\mathbf{P}^*) \le -\frac{\|\mathbf{s}-\mathbf{P}^*\|_2^2}{2 (N_{\max} +P)^2 }. \label{eqnTaylor2nd}
\end{align}
For the sake of completeness, the derivations of~\eqref{eqnTaylor1st} and~\eqref{eqnTaylor2nd} are contained in Appendix~\ref{appendixB}.
Combining~\eqref{eqnTaylor}, \eqref{eqnTaylor1st} and~\eqref{eqnTaylor2nd}, we have for all $\mathbf{s}\in \mathcal{S}^{(n)}$
\begin{align}
f(\mathbf{P}^*)-f(\mathbf{s}) \ge \frac{\|\mathbf{s}-\mathbf{P}^*\|_2^2}{4 (N_{\max} +P)^2 },
\end{align}
which together with the definitions of~$f$ and~$\mu_\ell$ in~\eqref{defFunctionF} and~\eqref{defMu} respectively implies that
\begin{align}
\mathrm{C}(\mathbf{P}^*)-\mathrm{C}(\mathbf{s}) &\ge \frac{\|\mathbf{s}-\mathbf{P}^*\|_2^2}{4 (N_{\max} +P)^2 } + \sum_{\ell=1}^L\mu_\ell(s_\ell-P_\ell) \\*
& \ge  \frac{\|\mathbf{s}-\mathbf{P}^*\|_2^2}{4 (N_{\max} +P)^2 }.
\end{align}
Consequently, \eqref{eqnBHT6thChain} holds by setting
\begin{equation}
\kappa\stackrel{\text{def}}{=} \frac{1}{4 (N_{\max} +P)^2}. \label{defKappa}
\end{equation}
Following~\eqref{eqnBHT5th2Chain}, we consider for each $\mathbf{s}\in \mathcal{S}^{(n)}\setminus\Pi^{(n)}$
\begin{align}
&\Pr_{p_{\boldsymbol{X}^n, \boldsymbol{Y}^n}^{(\mathbf{s})}}\left\{\frac{1}{\sqrt{n}}\sum_{k=1}^n U_k^{(\mathbf{s})}\ge \sqrt{n}\big(\mathrm{C}(\mathbf{P}^*)-\mathrm{C}(\mathbf{s})\big)+ \sqrt{\mathrm{V}(\mathbf{P}^*)}\, \Phi^{-1}(\varepsilon+\tau)\right\}\notag\\
&\le \Pr_{p_{\boldsymbol{X}^n, \boldsymbol{Y}^n}^{(\mathbf{s})}}\left\{\frac{1}{\sqrt{n}}\sum_{k=1}^n U_k^{(\mathbf{s})}\ge \kappa\sqrt{n}\|\mathbf{P}^*-\mathbf{s}\|_{2}^2+ \sqrt{\mathrm{V}(\mathbf{P}^*)}\, \Phi^{-1}(\varepsilon+\tau)\right\} \label{eqnBHT7thChaina}\\
&\le \Pr_{p_{\boldsymbol{X}^n, \boldsymbol{Y}^n}^{(\mathbf{s})}}\left\{\frac{1}{\sqrt{n}}\sum_{k=1}^n U_k^{(\mathbf{s})}\ge \kappa n^{1/6}+ \sqrt{\mathrm{V}(\mathbf{P}^*)}\, \Phi^{-1}(\varepsilon+\tau)\right\} \label{eqnBHT7thChainb}
\end{align}
where
\begin{itemize}
\item \eqref{eqnBHT7thChaina} is due to~\eqref{eqnBHT6thChain}.
\item \eqref{eqnBHT7thChainb} follows from the definition of~$\Pi^{(n)}$ in~\eqref{defSetPiN}.
\end{itemize}
Following the standard approach for obtaining large deviation bounds, we apply Markov's inequality on the RHS of~\eqref{eqnBHT7thChainb} and obtain for each $\mathbf{s}\in \mathcal{S}^{(n)}\setminus\Pi^{(n)}$
\begin{align}
&\Pr_{p_{\boldsymbol{X}^n, \boldsymbol{Y}^n}^{(\mathbf{s})}}\left\{\frac{1}{\sqrt{n}}\sum_{k=1}^n U_k^{(\mathbf{s})}\ge \sqrt{n}\big(\mathrm{C}(\mathbf{P}^*)-\mathrm{C}(\mathbf{s})\big)+ \sqrt{\mathrm{V}(\mathbf{P}^*)}\, \Phi^{-1}(\varepsilon+\tau)\right\}\notag\\
&\le \frac{\E_{p_{\boldsymbol{X}^n, \boldsymbol{Y}^n}^{(\mathbf{s})}}\left[e^{\frac{1}{\sqrt{n}}\sum_{k=1}^n U_k^{(\mathbf{s})}}\right]}{e^{\kappa n^{1/6}+ \sqrt{\mathrm{V}(\mathbf{P}^*)}\, \Phi^{-1}(\varepsilon+\tau)}}. \label{eqnBHT8thChain}
\end{align}
In order to bound the RHS of~\eqref{eqnBHT8thChain}, consider the following chain of inequalities for each $\mathbf{s}\in \mathcal{S}^{(n)}\setminus\Pi^{(n)}$:
\begin{align}
\E_{p_{\boldsymbol{X}^n, \boldsymbol{Y}^n}^{(\mathbf{s})}}\left[e^{\frac{1}{\sqrt{n}}\sum_{k=1}^n U_k^{(\mathbf{s})}}\right]& = \left(\prod_{\ell=1}^L\frac{s_\ell+N_\ell}{(1+n^{-1/2})s_\ell+N_\ell}\right)^{n/2}e^{\sum\limits_{\ell=1}^L\left(\frac{\sqrt{n}s_\ell}{2(s_\ell+N_\ell)}+\frac{N_\ell s_\ell}{2(s_\ell+N_\ell)\left(\left(1+n^{-1/2}\right)s_\ell+N_\ell\right)}\right)}\label{eqnBHT9thChaina}\\
  &\le \left(\prod_{\ell=1}^L\left(1-\frac{n^{-1/2}s_\ell}{(1+n^{-1/2})s_\ell+N_\ell}\right)^{n/2}e^{\frac{\sqrt{n}s_\ell}{2(s_\ell+N_\ell)}}\right)e^{\sum\limits_{\ell=1}^L\frac{N_\ell s_\ell}{2(s_\ell+N_\ell)^2}} \\
& \le e^{\sum\limits_{\ell=1}^L\left( \frac{s_\ell^2}{2((1+n^{-1/2})s_\ell+N_\ell)(s_\ell+N_\ell)} +\frac{N_\ell s_\ell}{2(s_\ell+N_\ell)^2}\right)} \label{eqnBHT9thChainb}\\
&\le e^{\sum\limits_{\ell=1}^L\frac{ s_\ell}{2(s_\ell+N_\ell)}} \\*
&\le e^{L/2},\label{eqnBHT9thChain}
\end{align}
where
\begin{itemize}
\item \eqref{eqnBHT9thChaina} follows from straightforward calculations based on the definition of $U_k^{(\mathbf{s})}$ in~\eqref{defUkd}, the property of~$p_{\boldsymbol{X}^n, \boldsymbol{Y}^n}^{(\mathbf{s})}$ in~\eqref{defDistPs} and the channel law, which are elaborated in Appendix~\ref{appendixD} for the sake of completeness.
\item \eqref{eqnBHT9thChainb} is due to the fact that $(1-\frac{1}{x})^x \le e^{-1}$ for all $x>1$.
\end{itemize}
Combining~\eqref{eqnBHT8thChain} and~\eqref{eqnBHT9thChain}, we have the following large deviation bound for each $\mathbf{s}\in \mathcal{S}^{(n)}\setminus\Pi^{(n)}$:
\begin{align}
\Pr_{p_{\boldsymbol{X}^n, \boldsymbol{Y}^n}^{(\mathbf{s})}}\left\{\frac{1}{\sqrt{n}}\sum_{k=1}^n U_k^{(\mathbf{s})}\ge \sqrt{n}\big(\mathrm{C}(\mathbf{P}^*)-\mathrm{C}(\mathbf{s})\big)+ \sqrt{\mathrm{V}(\mathbf{P}^*)}\, \Phi^{-1}(\varepsilon+\tau)\right\}\le\frac{e^{L/2}}{e^{\kappa n^{1/6}+ \sqrt{\mathrm{V}(\mathbf{P}^*)}\, \Phi^{-1}(\varepsilon+\tau)}}. \label{eqnBHT9*thChain}
\end{align}
Following~\eqref{eqnBHT5th2Chain}, we use \eqref{eqnBHT9*thChain} and~\eqref{cardinalityBound} to obtain
\begin{align}
\lim_{n\rightarrow\infty}\sum_{\mathbf{s}\in \mathcal{S}^{(n)}\setminus\Pi^{(n)}}\Pr_{p_{\boldsymbol{X}^n, \boldsymbol{Y}^n}^{(\mathbf{s})}}\left\{\frac{1}{\sqrt{n}}\sum_{k=1}^n U_k^{(\mathbf{s})}\ge \sqrt{n}\big(\mathrm{C}(\mathbf{P}^*)-\mathrm{C}(\mathbf{s})\big)+ \sqrt{\mathrm{V}(\mathbf{P}^*)}\, \Phi^{-1}(\varepsilon+\tau)\right\}=0. \label{eqnBHT10thChain}
\end{align}
\subsection{Combining Earlier Bounds to Obtain an Upper Bound on~$\mathrm{L}_\varepsilon^{\text{fb}}$}
Combining~\eqref{eqnBHTSecondChain}, \eqref{defxin}, \eqref{eqnBHT5th2Chain}, \eqref{levyThmEq4}, \eqref{eqnBHT10thChain} and~\eqref{cardinalityBound}, we have
\begin{align}
\log M_{\mathrm{fb}}^*(\bar n, \varepsilon, P)
 \le n\mathrm{C}(\mathbf{P}^*) + \sqrt{n}\left(\sqrt{\mathrm{V}(\mathbf{P}^*)}\, \Phi^{-1}(\varepsilon+\tau)\right) + \log\big(2n^L\big) - \log\left(\tau/2\right)  \label{eqnBHT11thChain}
\end{align}
for all sufficiently large~$n$, which together with~\eqref{defBarN} implies that
\begin{equation}
\liminf\limits_{\bar n\rightarrow \infty}\frac{1}{\sqrt{\bar n}}\big(\log M_{\mathrm{fb}}^*(\bar n, \varepsilon, P)-\bar n\mathrm{C}(\mathbf{P}^*)\big) \le \sqrt{\mathrm{V}(\mathbf{P}^*)}\, \Phi^{-1}(\varepsilon+ \tau). \label{eqnBHT12thChain}
\end{equation}
Since $\tau>0$ is arbitrary, it follows from~\eqref{eqnBHT12thChain} and Definition~\ref{defDispersion} that
\begin{equation}
\mathrm{L}_\varepsilon^{\text{fb}}\le \sqrt{\mathrm{V}(\mathbf{P}^*)}\, \Phi^{-1}(\varepsilon).
\end{equation}
\section{Concluding Remarks} \label{sectionConclusion}
\subsection{Novel Ingredients in Proof of Theorem~\ref{thmMainResult}} \label{sectionNovelIngredients}
As mentioned in Section~\ref{subsecRelatedWork}, the proof of~\cite[Th.~ 2]{AW14} which obtains upper bounds on the second-order asymptotics of DMCs with feedback cannot be generalized to the parallel Gaussian channel with feedback. Indeed, the proof of Theorem~\ref{thmMainResult} follows the standard procedures for obtaining the second-order asymptotics of DMCs without feedback (see, e.g., \cite[proof of Th.~50]{Pol10} and~\cite[Sec.~III]{TomTan12}) except the following three novel ingredients:
\begin{enumerate}
\item Instead of classifying transmitted codewords into polynomially many type classes based on their empirical distributions which is generally not possible for channels with continuous input alphabet, we discretize the transmitted power and classify the codewords into polynomially many type classes based on their discretized power types. In particular, the collection of \emph{power type classes} $\mathcal{S}^{(n)}$ in~\eqref{defSetS} plays a key role in our analysis, and there are polynomially many power type classes by~\eqref{cardinalityBound}.
    The details can be found in Section~\ref{stepAinProof} in the proof.
\item Curtiss' theorem rather than Berry-Ess\'een theorem is invoked to bound the information spectrum term (the first term in~\eqref{eqnBHT5th2Chain}) related to transmitted codewords whose types are close to the optimal power allocation. In particular, Berry-Ess\'een theorem for bounded martingale difference sequences cannot be used to bound the information spectrum term in the presence of feedback because each input symbol~$X_{\ell,k}$ belongs to an interval $[-\sqrt{nP}, \sqrt{nP}]$ that grows unbounded as~$n$ increases. Instead, we apply Curtiss' theorem to show that the distribution of the sum of random variables in the information spectrum term converges to some distribution generated from a sum of i.i.d.\ random variables (i.e., \eqref{levyThmEq1}), thus facilitating the use of the usual central limit theorem~\cite{feller}. The details can be found in Section~\ref{stepBInProof}.
    \item In order to bound the information spectrum term related to transmitted codewords whose types are far from the optimal power allocation (the second term in~\eqref{eqnBHT5th2Chain}), the usual approach is to bound the information spectrum term by an \emph{average } of exponentially many upper bounds where each upper bound is then further simplified by invoking Chebyshev's inequality~\cite[Sec.~X-A]{Hayashi09}. However, due to the presence of feedback, the information spectrum term can be expressed as only a sum (instead of average) of polynomially many upper bounds as shown in the second term in~\eqref{eqnBHT5th2Chain}. In order to control the \emph{sum} of polynomially many upper bounds, we have to resort to large deviation bounds as shown in~\eqref{eqnBHT9*thChain} rather than the weaker Chebyshev's inequality. The details can be found in Section~\ref{stepCInProof}.
\end{enumerate}

\subsection{Major Difficulties in Tightening the Third-Order Term}
If the feedback link is absent, the third-order term of the optimal finite blocklenth rate $\frac{1}{n}\log M_{\text{fb}}^*(n,\varepsilon,P)$ is $\Theta\Big(\frac{\log n}{n}\Big)$ as shown in~\eqref{eqn:asymp_expans} in Section~\ref{Introduction}.
The third-order term can be obtained by applying Berry-Ess\'een theorem to bound an information spectrum term (analogous to the first term in~\eqref{eqnBHT5th2Chain}).

In the presence of feedback, Theorem~\ref{thmMainResult} shows that the third-order term is $o\Big(\frac{1}{\sqrt{n}}\Big)$. If we want to compute an explicit upper bound on the third-order term using the current proof technique, an intuitive way is to invoke a non-asymptotic version of Curtiss' theorem that can measure the proximity between two distributions based on the proximity between their moment generating functions. However, such a non-asymptotic version of Curtiss' theorem does not exist to the best of our knowledge, which prohibits us from strengthening the current $o\Big(\frac{1}{\sqrt{n}}\Big)$ bound on the third-order term.
It is worth noting that~\eqref{levyThmEq1*} and~\eqref{levyThmEq2} in our proof break down if the moment generating functions are replaced with characteristic functions. If one can find a way to make characteristic functions amenable to our proof approach, then
a non-asymptotic version of L\'evy's continuity theorem known as \emph{Ess\'een's smoothing lemma} (see, e.g., \cite[Th.~1.5.2]{IbragimovLinnik1971}) may be invoked to tighten the third-order term herein.
\subsection{Future Work} \label{sectionFutureWork}
The techniques presented herein may be used to analyze the fixed-error asymptotics of fixed-length codes over parallel DMCs with cost constraint and with feedback. We envision that there will be an added layer of complexity as the {\em method of types}~\cite{Csi97} is typically used to analyze the fixed-error asymptotics of DMCs with and without cost constraint~\cite{Strassen}. Hence, we anticipate that {\em two} applications of the method of types need to be applied --- one for handling power types that specify the power allocation among the parallel channels (as was done in Section~\ref{stepAinProof}) and another for handling codewords of the same power type but of different empirical distributions (the usual types).
 In the present work, the latter situation is ameliorated by the fact that the maximum rate achievable by a coding scheme over an AWGN channel is completely determined by the power allocated for the coding scheme.
\appendices
\section{Proof of Lemma~\ref{lemmaTransformedCode}}\label{appendixA}
\begin{IEEEproof}
Let $f_{\ell,k}$ and $\tilde f_{\ell,k}$ be the encoding functions of the $(n, M, P)$-feedback code and the $(\gamma, \mathbf{s})$-modified code respectively for each $\ell\in\mathcal{L}$ and each $k\in\{1, 2, \ldots, n\}$. For any $w\in\mathcal{W}$ and any $\mathbf{y}^n\in \mathbb{R}^{L\times n}$ such that
\begin{align}
\phi\left( \left[\begin{matrix}
f_{1,1}(w) \\\vdots\\ f_{L,1}(w)
\end{matrix} \right]
,\ldots, \left[\begin{matrix}
f_{1,n}(w, \mathbf{y}^{n-1}) \\\vdots\\ f_{L,n}(w, \mathbf{y}^{n-1})
\end{matrix} \right]\right)\in \Gamma^{(\gamma)}(\mathbf{s}) \label{lemmaTransformedCodeEq1}
\end{align}
and
\begin{align}
\sum_{\ell=1}^L\sum_{k=1}^n f_{\ell,k}(w, \mathbf{y}^{k-1})^2=nP, \label{lemmaTransformedCodeEq2}
\end{align}
it follows from~\eqref{defTildeFeq1}, \eqref{defTildeFeq2} and~\eqref{defTildeFeq3} in Definition~\ref{definitionTransformedCode}
 that
\begin{align}
\left( \left[\begin{matrix}
f_{1,1}(w) \\\vdots\\ f_{L,1}(w)
\end{matrix} \right],\ldots,
\left[\begin{matrix}
f_{1,n}(w, \mathbf{y}^{n-1}) \\\vdots\\ f_{L,n}(w, \mathbf{y}^{n-1})
\end{matrix} \right]\right)=
\left( \left[\begin{matrix}
\tilde f_{1,1}(w) \\\vdots\\ \tilde f_{L,1}(w)
\end{matrix} \right]
,\ldots, \left[\begin{matrix}
\tilde f_{1,n}(w, \mathbf{y}^{n-1}) \\\vdots\\ \tilde f_{L,n}(w, \mathbf{y}^{n-1})
\end{matrix} \right]\right). \label{lemmaTransformedCodeEq3}
\end{align}
Since~\eqref{lemmaTransformedCodeEq3} holds for any $w\in\mathcal{W}$ and $\mathbf{y}^n\in \mathbb{R}^{L\times n}$ that satisfy~\eqref{lemmaTransformedCodeEq1} and~\eqref{lemmaTransformedCodeEq2}, it follows that~\eqref{lemmaTransformedCodeSt} holds for all Borel measurable~$\mathcal{A}\subseteq\mathbb{R}^{L\times n}\times \mathbb{R}^{L\times n}$.
\end{IEEEproof}

\section{Proof of Lemma~\ref{lemma1InProof}}\label{appendixC}
The proof for the AWGN channel case (i.e., $L=1$) is contained in~\cite[Sec.~E]{FongTan15}. For a general~$L\in \mathbb{N}$, consider the following chain of equalities for each $m=n, n-1, \ldots, 1$:
 \begin{align}
& \E\left[e^{\lambda \sum\limits_{k=1}^m U_k^{(\mathbf{P}^*)}}\cdot e^{\lambda ^2\sum\limits_{\ell=1}^L\frac{N_\ell \left(nP_\ell  -\sum\limits_{k=1}^m X_{\ell,k}^2\right) }{2(P_\ell+N_\ell)\left((1+\lambda )P_\ell+N_\ell\right)}}\right] \notag\\
& =  \E\left[\E\left[\left.e^{\lambda\sum\limits_{k=1}^{m} U_k^{(\mathbf{P}^*)}}\cdot e^{\lambda^2\sum\limits_{\ell=1}^L\frac{N_\ell \left(nP_\ell  -\sum\limits_{k=1}^{m} X_{\ell,k}^2\right) }{2(P_\ell+N_\ell)\left((1+\lambda  )P_\ell+N_\ell\right)}}\right|\boldsymbol{X}^{m}, \boldsymbol{Z}^{m-1}\right]\right] \\
& = \E\left[ e^{\lambda\sum\limits_{k=1}^{m-1} U_k^{(\mathbf{P}^*)}}\cdot e^{\lambda^2\sum\limits_{\ell=1}^L\frac{N_\ell \left(nP_\ell  -\sum\limits_{k=1}^{m-1} X_{\ell,k}^2\right) }{2(P_\ell+N_\ell)\left((1+\lambda  )P_\ell+N_\ell\right)}} \cdot \int_{\mathbb{R}^{L\times n}}p_{\boldsymbol{Z}_m}^*(\mathbf{z}_m) \cdot e^{\lambda U_m^{(\mathbf{P}^*)}}\cdot e^{\lambda^2\sum\limits_{\ell=1}^L \frac{- X_{\ell,m}^2 }{2(P_\ell+N_\ell)\left((1+\lambda  )P_\ell+N_\ell\right)}} \mathrm{d}\mathbf{z}_m \right] \label{appendixCeq2b-}\\
&= e^{\lambda\sum\limits_{\ell=1}^L\frac{P_\ell}{2(P_\ell+N_\ell)}}\sqrt{\prod_{\ell=1}^L\frac{P_\ell+N_\ell}{(1+\lambda  )P_\ell+N_\ell}} \cdot \E\left[e^{\lambda\sum\limits_{k=1}^{m-1} U_k^{(\mathbf{P}^*)}}\cdot e^{\lambda^2\sum\limits_{\ell=1}^L\frac{N_\ell \left(nP_\ell  -\sum\limits_{k=1}^{m-1} X_{\ell,k}^2\right) }{2(P_\ell+N_\ell)\left((1+\lambda  )P_\ell+N_\ell\right)}}\right] \label{appendixCeq2b}
 \end{align}
 where 
 \begin{itemize}
 \item  \eqref{appendixCeq2b-} is due to the fact that $\boldsymbol{Z}_m$ and $(\boldsymbol{X}^{m},\boldsymbol{Z}^{m-1})$ are independent.
     \item \eqref{appendixCeq2b} is due to the definition of~$U_k^{(\mathbf{P}^*)}$ in~\eqref{defUkd}.
 \end{itemize}
 Applying \eqref{appendixCeq2b} recursively from $m=n$ to $m=1$, we have
  \begin{align}
  & \E\left[e^{\lambda\sum\limits_{k=1}^n U_k^{(\mathbf{P}^*)}}\cdot e^{\lambda^2\sum\limits_{\ell=1}^L\frac{N_\ell \left(nP_\ell  -\sum\limits_{k=1}^n X_{\ell,k}^2\right) }{2(P_\ell+N_\ell)\left((1+\lambda  )P_\ell+N_\ell\right)}}\right] \notag\\
& =\left(\prod_{\ell=1}^L\frac{P_\ell+N_\ell}{(1+\lambda  )P_\ell+N_\ell}\right)^{n/2}e^{n\sum\limits_{\ell=1}^L\left(\frac{\lambda P_\ell}{2(P_\ell+N_\ell)}+\frac{\lambda^2N_\ell P_\ell}{2(P_\ell+N_\ell)\left((1+\lambda  )P_\ell+N_\ell\right)}\right)}. \label{appendixCeq3}
  \end{align}
  On the other hand, straightforward calculations based on  the definition of~$V_k^{(\mathbf{P}^*)}$ in~\eqref{defVkd} and the fact that $\{\boldsymbol{Z}_k\}_{k=1}^n$ are independent implies that
 \begin{align}
 \E\left[e^{\lambda\sum\limits_{k=1}^n V_k^{(\mathbf{P}^*)}}\right]
& =\left(\prod_{\ell=1}^L\frac{P_\ell+N_\ell}{(1+\lambda  )P_\ell+N_\ell}\right)^{n/2}e^{n\sum\limits_{\ell=1}^L\left(\frac{\lambda P_\ell}{2(P_\ell+N_\ell)}+\frac{\lambda^2N_\ell P_\ell}{2(P_\ell+N_\ell)\left((1+\lambda  )P_\ell+N_\ell\right)}\right)}. \label{appendixCeq4}
 \end{align}
 Combining~\eqref{appendixCeq3} and~\eqref{appendixCeq4}, we obtain~\eqref{appendixCeq1}.

 \section{Derivations of~\eqref{eqnTaylor1st} and~\eqref{eqnTaylor2nd}} \label{appendixB}
Straightforward calculations based on~\eqref{defFunctionF} reveal that for all $\mathbf{s}\in \mathbb{R}_+^L$, we obtain that
\begin{align}
\triangledown f(\mathbf{s}) = \frac{1}{2}\left[\begin{matrix}\frac{1}{N_1 + s_1}-\frac{1}{2\Lambda}+\mu_1 \\ \vdots\\  \frac{1}{N_L + s_L}-\frac{1}{2\Lambda}+\mu_L \end{matrix}\right] \label{appendixBeq1}
\end{align}
and $\triangledown^2 f(\mathbf{s})$ is a diagonal matrix that satisfies
\begin{align}
\triangledown^2 f(\mathbf{s}) = \frac{-1}{2}\left[\begin{matrix}\frac{1}{(N_1 + s_1)^2} & 0 & \ldots & 0\\ 0& \frac{1}{(N_2 + s_2)^2} &\ldots & 0\\\vdots & \vdots & \ddots &  \vdots\\0&0&\ldots& \frac{1}{(N_L + s_L)^2}\end{matrix}\right].  \label{appendixBeq2}
\end{align}
Combining~\eqref{appendixBeq1}, \eqref{sumPell=P}, \eqref{pEllValue} and~\eqref{defMu}, we have~$\triangledown f(\mathbf{P}^*) = 0$. In addition, for any $\mathbf{s}$ such that $\sum_{\ell=1}^L s_\ell \le P$, it follows from~\eqref{appendixBeq2} that $N_\ell + s_\ell \le N_{\max} +P$ for all $\ell\in\mathcal{L}$, which then implies that~\eqref{eqnTaylor2nd} holds for all $\mathbf{s}\in \mathcal{S}^{(n)}$.

\section{Derivation of~\eqref{eqnBHT9thChaina}}\label{appendixD}
Let $t=\frac{1}{\sqrt{n}}$. Fix any $\mathbf{s}\in \mathcal{S}^{(n)}\setminus\Pi^{(n)}$. Due to~\eqref{defDistPs}, it suffices to show that
 \begin{align}
&\E_{p_{\boldsymbol{X}^n, \boldsymbol{Y}^n}^{(\mathbf{s})}}\left[e^{t\sum\limits_{k=1}^n U_k^{(\mathbf{s})}}\cdot e^{t^2\sum\limits_{\ell=1}^L\frac{N_\ell \left(ns_\ell  -\sum\limits_{k=1}^n X_{\ell,k}^2\right) }{2(s_\ell+N_\ell)\left((1+t)s_\ell+N_\ell\right)}}\right] \notag\\
 &= \left(\prod_{\ell=1}^L\frac{s_\ell+N_\ell}{(1+t)s_\ell+N_\ell}\right)^{n/2}e^{n\sum\limits_{\ell=1}^L\left(\frac{ts_\ell}{2(s_\ell+N_\ell)}+\frac{t^2N_\ell s_\ell}{2(s_\ell+N_\ell)\left((1+t)s_\ell+N_\ell\right)}\right)}. \label{appendixDeq1}
 \end{align}
Replacing $\mathbf{P}^*$ with $\mathbf{s}$ in the steps leading to~\eqref{appendixCeq2b} and~\eqref{appendixCeq3}, we obtain~\eqref{appendixDeq1}.



\ifCLASSOPTIONcaptionsoff
 \newpage
\fi




\begin{thebibliography}{10}
\providecommand{\url}[1]{#1}
\csname url@samestyle\endcsname
\providecommand{\newblock}{\relax}
\providecommand{\bibinfo}[2]{#2}
\providecommand{\BIBentrySTDinterwordspacing}{\spaceskip=0pt\relax}
\providecommand{\BIBentryALTinterwordstretchfactor}{4}
\providecommand{\BIBentryALTinterwordspacing}{\spaceskip=\fontdimen2\font plus
\BIBentryALTinterwordstretchfactor\fontdimen3\font minus
  \fontdimen4\font\relax}
\providecommand{\BIBforeignlanguage}[2]{{%
\expandafter\ifx\csname l@#1\endcsname\relax
\typeout{** WARNING: IEEEtran.bst: No hyphenation pattern has been}%
\typeout{** loaded for the language `#1'. Using the pattern for}%
\typeout{** the default language instead.}%
\else
\language=\csname l@#1\endcsname
\fi
#2}}
\providecommand{\BIBdecl}{\relax}
\BIBdecl

\bibitem{CoverBook}
{T. M. Cover and J. A. Thomas}, \emph{{Elements of Information Theory}},
  2nd~ed.\hskip 1em plus 0.5em minus 0.4em\relax Hoboken, NJ: John Wiley and
  Sons, 2006.

\bibitem{elgamal}
A.~{El~Gamal} and Y.-H. Kim, \emph{Network Information Theory}.\hskip 1em plus
  0.5em minus 0.4em\relax Cambridge, U.K.: Cambridge University Press, 2012.

\bibitem{davidTseBook}
D.~Tse and P.~Viswanath, \emph{Fundamentals of Wireless Communication}.\hskip
  1em plus 0.5em minus 0.4em\relax Cambridge, U.K.: Cambridge University Press,
  2005.

\bibitem{Sha49}
C.~E. Shannon, ``Communication in the presence of noise,'' \emph{Proceedings of
  IRE}, vol.~37, no.~1, pp. 10--21, 1949.

\bibitem{Pol10}
Y.~Polyanskiy, ``Channel coding: {Non}-asymptotic fundamental limits,'' Ph.D.
  dissertation, Princeton University, 2010.

\bibitem{TanTom13a}
V.~Y.~F. Tan and M.~Tomamichel, ``The third-order term in the normal
  approximation for the {AWGN} channel,'' \emph{{IEEE} Trans. Inf. Theory},
  vol.~61, no.~5, pp. 2430--2438, 2015.

\bibitem{Sha56}
C.~E. Shannon, ``The zero error capacity of a noisy channel,'' \emph{IRE
  Transactions on Information Theory}, vol.~2, no.~3, pp. 8--19, 1956.

\bibitem{AW14}
Y.~Altu\u{g} and A.~B. Wagner, ``Feedback can improve the second-order coding
  performance in discrete memoryless channels,'' in \emph{Proc. IEEE Intl.
  Symp. Inf. Theory}, Honolulu, HI, Jul 2014, pp. 2361--2365.

\bibitem{PPV11b}
Y.~Polyanskiy, H.~V. Poor, and S.~Verd\'{u}, ``Feedback in the non-asymptotic
  regime,'' \emph{{IEEE} Trans. Inf. Theory}, vol.~57, no.~8, pp. 4903--4925,
  2011.

\bibitem{Machkouri}
M.~E. Machkouri and L.~Ouchti, ``Exact convergence rates in the central limit
  theorem for a class of martingales,'' \emph{Bernoulli}, vol.~13, no.~4, pp.
  981--999, Nov. 2007.

\bibitem{feller}
W.~Feller, \emph{An Introduction to Probability Theory and Its Applications},
  2nd~ed.\hskip 1em plus 0.5em minus 0.4em\relax Hoboken, NJ: John Wiley and
  Sons, 1971, vol.~2.

\bibitem{FongTan15}
{S.~L.~Fong and V.~Y.~F.~Tan}, ``{Asymptotic expansions for the {AWGN} channel
  with feedback under a peak power constraint},'' in \emph{Proc. IEEE Intl.
  Symp. Inf. Theory}, Hong Kong, Jun. 2015, pp. 311--315.

\bibitem{TFT17}
L.~V. Truong, S.~L. Fong, and V.~Y.~F. Tan, ``On {Gaussian} channels with
  feedback under expected power constraints and with non-vanishing error
  probabilities,'' \emph{{IEEE} Trans. Inf. Theory}, vol.~63, no.~3, pp.
  1746--1765, 2017.

\bibitem{Wang09}
L.~Wang, R.~Colbeck, and R.~Renner, ``Simple channel coding bounds,'' in
  \emph{Proc. IEEE Intl. Symp. Inf. Theory}, Seoul, South Korea, June/July
  2009, pp. 1804 -- 1808.

\bibitem{PPV10}
Y.~Polyanskiy, H.~V. Poor, and S.~Verd\'{u}, ``Channel coding rate in the
  finite blocklength regime,'' \emph{{IEEE} Trans. Inf. Theory}, vol.~56,
  no.~5, pp. 2307--2359, 2010.

\bibitem{curtiss}
J.~H. Curtiss, ``A note on the theory of moment generating functions,''
  \emph{Ann. Math. Statist.}, vol.~13, no.~4, pp. 430--433, 1942.

\bibitem{Williams1991}
D.~Williams, \emph{Probability with Martingales}.\hskip 1em plus 0.5em minus
  0.4em\relax Cambridge, U.K.: Cambridge University Press, 1991.

\bibitem{Hayashi09}
M.~Hayashi, ``Information spectrum approach to second-order coding rate in
  channel coding,'' \emph{{IEEE} Trans. Inf. Theory}, vol.~55, no.~11, pp.
  4947--4966, 2009.

\bibitem{TomTan12}
M.~Tomamichel and V.~Y.~F. Tan, ``A tight upper bound for the third-order
  asymptotics of most discrete memoryless channels,'' \emph{{IEEE} Trans. Inf.
  Theory}, vol.~59, no.~11, pp. 7041--7051, 2013.

\bibitem{IbragimovLinnik1971}
I.~A. Ibragimov and Y.~V. Linnik, \emph{Independent and stationary sequences of
  random variables}, J.~F.~C. Kingman, Ed.\hskip 1em plus 0.5em minus
  0.4em\relax Groningen, Netherlands: Wolters-Noordhoff Publishing, 1971.

\bibitem{Csi97}
I.~Csisz\'{a}r and J.~{K\"{o}rner}, \emph{Information Theory: Coding Theorems
  for Discrete Memoryless Systems}.\hskip 1em plus 0.5em minus 0.4em\relax
  Cambridge, U.K.: Cambridge University Press, 2011.

\bibitem{Strassen}
V.~Strassen, ``{Asymptotische Absch\"{a}tzungen in Shannons
  Informationstheorie},'' in \emph{Trans. Third Prague Conf. Inf. Theory},
  Prague, 1962, pp. 689--723,
  http://www.math.cornell.edu/$\sim$pmlut/strassen.pdf.

\end{thebibliography}
%
%
%

\section*{Acknowledgments}
The authors would like to thank the Associate Editor Prof.\ Shun Watanabe and the two anonymous reviewers for the useful comments that improve the presentation of this paper.
\end{document}